\documentclass[11pt,a4paper]{article}

\usepackage{amssymb}
\usepackage[dvips]{graphicx}
\usepackage{bm}

\begin{document}

\title{\bf Two-loop renormalization of the Faddeev--Popov ghosts in ${\cal N}=1$ supersymmetric gauge theories regularized by higher derivatives}

\author{
\smallskip
A.E.Kazantsev, M.D.Kuzmichev, N.P.Meshcheriakov, S.V.Novgorodtsev,\\
\medskip
I.E.Shirokov, M.B.Skoptsov, K.V.Stepanyantz\\
{\small{\em Moscow State University}}, {\small{\em  Physical
Faculty, Department  of Theoretical Physics}}\\
{\small{\em 119991, Moscow, Russia}}}

\maketitle

\begin{abstract}
For the general renormalizable ${\cal N}=1$ supersymmetric gauge theory we investigate renormalization of the Faddeev--Popov ghosts using the higher covariant derivative regularization. First, we find the two-loop anomalous dimension defined in terms of the bare coupling constant in the general $\xi$-gauge. It is demonstrated that for doing this calculation one should take into account that the quantum gauge superfield is renormalized in a nonlinear way. Next, we obtain the two-loop anomalous dimension of the Faddeev--Popov ghosts defined in terms of the renormalized coupling constant and examine its dependence on the subtraction scheme.
\end{abstract}

\unitlength=1cm

\section{Introduction}
\hspace*{\parindent}

Renormalization of ${\cal N}=1$ supersymmetric theories has a lot of interesting features. For example, according to the well-known non-renormalization theorem \cite{Grisaru:1979wc} the superpotential has no divergent quantum corrections. Consequently, the renormalization of masses and Yukawa couplings is related to the renormalization of the chiral matter superfields. Also there are some other equations relating the renormalization group functions in ${\cal N}=1$ supersymmetric theories. In particular, the $\beta$-function can be expressed in terms of the anomalous dimension of the matter superfields $\big(\gamma_\phi\big)_i{}^j$ by the NSVZ equation \cite{Novikov:1983uc,Jones:1983ip,Novikov:1985rd,Shifman:1986zi} (see also Refs. \cite{Shifman:1999mv,Shifman:1999kf,Shifman:2018yxh} for a more recent discussion)

\begin{equation}\label{NSVZ_Original_Equation}
\beta(\alpha,\lambda) = - \frac{\alpha^2\Big(3 C_2 - T(R) + C(R)_i{}^j
\big(\gamma_\phi\big)_j{}^i(\alpha,\lambda)/r\Big)}{2\pi(1- C_2\alpha/2\pi)}.
\end{equation}

\noindent
Here $r$ denotes the dimension of the gauge group $G$ with the structure constants $f^{ABC}$, such that $f^{ACD} f^{BCD} \equiv C_2 \delta^{AB}$. The matter superfields transform according to a representation $R$. We denote the generators of this representation by $T^A$. They should be distinguished from the generators of the fundamental representation $t^A$. In our notation

\begin{equation}
\mbox{tr}(T^A T^B) = T(R)\delta^{AB};\qquad (T^A)_i{}^k (T^A)_k{}^j \equiv C(R)_i{}^j.
\end{equation}

\noindent
Eq. (\ref{NSVZ_Original_Equation}) is usually called the exact NSVZ $\beta$-function, because for the ${\cal N}=1$ supersymmetric Yang--Mills theory (SYM) without matter it really gives the exact expression for the $\beta$-function in the form of the geometric series. However, Eq. (\ref{NSVZ_Original_Equation}) is scheme-dependent \cite{Kutasov:2004xu,Kataev:2014gxa} and is valid only in certain (NSVZ) subtraction schemes. With dimensional reduction \cite{Siegel:1979wq,Siegel:1980qs} such schemes can be obtained from the $\overline{\mbox{DR}}$-scheme by finite renormalizations in each order of perturbation theory \cite{Jack:1996vg,Jack:1996cn,Jack:1998uj}. The all-order prescriptions giving the NSVZ scheme have been constructed in the case of using the Slavnov higher covariant derivative regularization \cite{Slavnov:1971aw,Slavnov:1972sq,Slavnov:1977zf} in the supersymmetric version \cite{Krivoshchekov:1978xg,West:1985jx}. It was done in Refs. \cite{Kataev:2014gxa,Kataev:2013eta,Kataev:2013csa} for the Abelian case and in Ref. \cite{Stepanyantz:2016gtk} for the non-Abelian case. Namely, to obtain the NSVZ scheme with the higher derivative regularization one should include into the renormalization constants only powers of $\ln\Lambda/\mu$, where $\Lambda$ is the dimensionful parameter of the regularized theory playing the role of the UV cut-off and $\mu$ is the renormalization scale. This prescription looks very similar to the one giving the MS scheme in the case of using the dimensional technique, when only $\varepsilon$-poles are included into renormalization constants. That is why the above described NSVZ scheme obtained with the higher derivative regularization can be called $\mbox{HD}+\mbox{MSL}$, where HD stands for higher derivatives and MSL means Minimal Subtraction of Logarithms, see, e.g. \cite{Kazantsev:2017fdc,Kataev:2017qvk,Stepanyantz:2017sqg}. This prescription has been verified by explicit three-loop calculations in the Abelian case in Refs. \cite{Kataev:2014gxa,Kataev:2013eta,Kataev:2013csa} and in the non-Abelian case for the terms containing Yukawa couplings in Refs. \cite{Shakhmanov:2017soc,Kazantsev:2018nbl}.

Note that the derivation of the NSVZ relation in the non-Abelian case by direct summation of supergraphs seems to involve the non-renormalization theorem for vertices with two legs of the Faddeev--Popov ghosts and one leg of the quantum gauge superfield \cite{Stepanyantz:2016gtk}. Finiteness of these vertices gives the relation between the renormalization constants of the coupling constant ($Z_\alpha$), of the Faddeev--Popov ghosts ($Z_c$), and of the quantum gauge superfield ($Z_V$)

\begin{equation}\label{Z_Relation}
\frac{d}{d\ln\Lambda} (Z_\alpha^{-1/2} Z_c Z_V) = 0.
\end{equation}

\noindent
Note that the quantum gauge superfield is renormalized in a nonlinear way \cite{Juer:1982fb,Juer:1982mp} according to the general arguments considered in \cite{Piguet:1981fb,Piguet:1981hh} (see also Refs. \cite{Piguet:1981mu,Piguet:1984mv}). In Eq. (\ref{Z_Relation}) we need only a coefficient of the linear term, although in this paper we will demonstrate that the nonlinear renormalization is very important for calculating the renormalization group functions.

Eq. (\ref{Z_Relation}) allows rewriting the NSVZ relation (\ref{NSVZ_Original_Equation}) in a different form

\begin{equation}\label{NSVZ_Equation_New}
\frac{\beta(\alpha,\lambda)}{\alpha^2} = - \frac{1}{2\pi}\Big(3 C_2 - T(R) - 2C_2 \gamma_c(\alpha,\lambda) - 2C_2 \gamma_V(\alpha,\lambda) + C(R)_i{}^j \big(\gamma_\phi\big)_j{}^i(\alpha,\lambda)/r\Big),
\end{equation}

\noindent
where $\gamma_c$ and $\gamma_V$ are the anomalous dimensions of the Faddeev--Popov ghosts and of the quantum gauge superfield, respectively. It is this form that follows from the perturbative calculations, see, e.g., the calculations in Refs. \cite{Kazantsev:2018nbl,Shakhmanov:2017wji}. As in the Abelian case, the NSVZ relation follows from the factorization of loop integrals into integrals of total \cite{Soloshenko:2003nc} and double total \cite{Smilga:2004zr} derivatives in the case of using the HD regularization.\footnote{Such a factorization does not take place for theories regularized by the dimensional reduction, see Refs. \cite{Aleshin:2015qqc,Aleshin:2016rrr}.} In the Abelian case this factorization was proved to be valid in all loops and to produce the NSVZ relation for the renormalization group functions defined in terms of the bare coupling constant \cite{Stepanyantz:2011jy,Stepanyantz:2014ima}. Similar arguments allow deriving the factorization into integrals of double total derivatives and the NSVZ-like relation for the Adler $D$-function \cite{Adler:1974gd} in ${\cal N}=1$ SQCD \cite{Shifman:2014cya,Shifman:2015doa}. The NSVZ-like relations existing in theories with softly broken supersymmetry \cite{Hisano:1997ua,Jack:1997pa,Avdeev:1997vx} can possibly be obtained by this method. For the anomalous dimension of the photino mass in softly broken ${\cal N}=1$ SQED it was demonstrated in \cite{Nartsev:2016nym}. As a consequence, in the cases mentioned above the NSVZ-like schemes are given by the $\mbox{HD}+\mbox{MSL}$ prescription in all loops \cite{Kataev:2017qvk,Nartsev:2016mvn}. For non-Abelian theories numerous calculations demonstrate similar features of quantum corrections as in the Abelian case \cite{Shakhmanov:2017soc,Aleshin:2016yvj,Pimenov:2009hv,Stepanyantz_MIAN,Stepanyantz:2011bz,Stepanyantz:2012zz,Stepanyantz:2012us,Kazantsev:2014yna}. However, at present there is no non-trivial verification of the term in Eq. (\ref{NSVZ_Equation_New}) which contains $\gamma_c$. Certainly, to obtain a non-trivial check, it is necessary to compare at least the three-loop contribution to the $\beta$-function with the two-loop contribution to the ghost anomalous dimension. For this purpose one has to know how the Faddeev--Popov ghosts are renormalized in the two-loop approximation in the case of using the HD regularization. Moreover, to verify Eq. (\ref{Z_Relation}) in the two-loop approximation (the one-loop check has been made in Ref. \cite{Aleshin:2016yvj}), the two-loop expression for $Z_c$ is also needed. That is why in this paper we calculate this renormalization constant in the two-loop approximation.

The paper is organized as follows: In Sect. \ref{Section_Theory} we briefly describe the theory under consideration and its regularization by higher derivatives. In the next Sect. \ref{Section_Anomalous_Dimension} we calculate the two-loop anomalous dimension of the Faddeev--Popov ghosts defined in terms of the bare coupling constant. The anomalous dimension defined in terms of the renormalized coupling constant is obtained in Sect. \ref{Section_Gamma_Renormalized}. The results are summarized in Conclusion.

\section{${\cal N}=1$ supersymmetric gauge theories with matter regularized by higher derivatives}
\hspace*{\parindent}\label{Section_Theory}

At the classical level the massless ${\cal N}=1$ supersymmetric gauge theory with matter is described by the action

\begin{eqnarray}\label{Action_Of_Theory}
&& S_{\mbox{\scriptsize classical}} = \frac{1}{2 e^2}\,\mbox{Re}\,\mbox{tr}\int d^4x\,
d^2\theta\,W^a W_a + \frac{1}{4} \int d^4x\, d^4\theta\,\phi^{*i}
(e^{2V})_i{}^j \phi_j\nonumber\\
&&\qquad\qquad\qquad\qquad\qquad\qquad\qquad\qquad\qquad +
\Big(\frac{1}{6}\lambda^{ijk} \int d^4x\,d^2\theta\, \phi_i \phi_j \phi_k + \mbox{c.c.}\Big),\qquad
\end{eqnarray}

\noindent
where we assume that the superpotential is cubic in the chiral matter superfields $\phi_i$ in order to obtain a renormalizable theory. In our notation the subscript $a$ numerates components of right spinors, while the subscripts with a dot (e.g., $\dot a$) numerate components of left spinors. The supersymmetric gauge field strength is described by the right Weyl spinor $W_a \equiv \bar D^2 (e^{-2V} D_a e^{2V})/8$, where $V$ is the Hermitian gauge superfield. Writing the action (\ref{Action_Of_Theory}) we assume that in the first term the gauge superfield inside $W_a$ is presented as
$V = e V^A t^A$, where $t^A$ are the generators of the fundamental representation, such that

\begin{equation}
\mbox{tr}(t^A t^B) = \frac{1}{2} \delta^{AB}; \qquad [t^A, t^B] = i f^{ABC} t^C.
\end{equation}

\noindent
If the matter superfields belong to a representation $R$ of the gauge group, then in the second term of Eq. (\ref{Action_Of_Theory}) $V = e V^A T^A$, where $T^A$ are the generators of the gauge group in the representation $R$, for which $\mbox{tr}(T^A T^B) = T(R) \delta^{AB}$.

Quantization of gauge theories in superspace allows obtaining a manifestly supersymmetric way of calculating quantum corrections. However, in the superfield formalism the renormalization of the theory (\ref{Action_Of_Theory}) is essentially different from the renormalization of the usual gauge theories. Namely, to absorb divergences, the gauge superfield $V$ should be renormalized in the nonlinear way \cite{Juer:1982fb,Juer:1982mp,Piguet:1981hh}.\footnote{Note that if quantum corrections in supersymmetric theories are calculated in components without eliminating auxiliary fields, then the auxiliary fields should also be renormalized non-linearly \cite{Jack:2005ij}.} Due to this nonlinear renormalization for quantization of the theory we should not only substitute the gauge and Yukawa couplings by the bare ones,

\begin{equation}
e \to e_0;\qquad \lambda^{ijk} \to \lambda_0^{ijk},
\end{equation}

\noindent
but also make the replacement $V\to {\cal F}(V)$, where ${\cal F}(V)$ is a nonlinear function which includes an infinite set of constants.\footnote{Note that now $V$ and $\phi_i$ also become the bare superfields.} This implies that $V = e_0 V^A t^A \to {\cal F}(V) = e_0 {\cal F}^A(V) t^A$ in the pure SYM part of the action (\ref{Action_Of_Theory}) and $V = e_0 V^A T^A \to {\cal F}(V) = e_0 {\cal F}^A(V) T^A$ in the matter part of the action. In the lowest-order approximation we can consider only terms cubic in the gauge superfield \cite{Juer:1982fb,Juer:1982mp},

\begin{equation}\label{Nonlinear_Function}
V^A \to {\cal F}^A(V) =  V^A + e_0^2\, y_0\, G^{ABCD} V^B V^C V^D + \ldots,
\end{equation}

\noindent
where dots denote terms of higher orders in $V$, and $y_0$ is a new bare constant. Due to the replacement (\ref{Nonlinear_Function}) the nonlinear renormalization in the lowest-order approximation is reduced to the linear renormalization of the constant $y_0$. The factor $e_0^2$ is included into the second term in order that the function ${\cal F}(V)$ should not contain the gauge coupling constant,

\begin{equation}
{\cal F}(V) = V + 8 y_0\, G^{ABCD}\, \mbox{tr}(V t^B)\, \mbox{tr}(V t^C)\, \mbox{tr}(V t^D)\, t^A + \ldots
\end{equation}

\noindent
According to \cite{Juer:1982fb,Juer:1982mp} the coefficient in lowest-order nonlinear term

\begin{equation}
G^{ABCD} = \frac{1}{6}\Big(f^{AKL} f^{BLM} f^{CMN} f^{DNK} + \mbox{permutations of $B$, $C$, and $D$} \Big)
\end{equation}

\noindent
is the totally symmetric tensor constructed from the structure constants. (It is easy to see that here the symmetrization with respect to the indices $BCD$ gives the totally symmetric expression with respect to the indices $ABCD$.)

The manifestly gauge invariant effective action can be constructed by the help of the background field method, which in the supersymmetric case is introduced by the substitution

\begin{equation}\label{Background_Substitution}
e^{2{\cal F}(V)} \to e^{\bm{\Omega}^+} e^{2{\cal F}(V)} e^{\bm{\Omega}}.
\end{equation}

\noindent
Then the background gauge superfield $\bm{V}$ is defined by the equation $e^{2\bm{V}} = e^{\bm{\Omega}^+} e^{\bm{\Omega}}$. After the replacement (\ref{Background_Substitution}) the action of the considered theory takes the form

\begin{eqnarray}\label{Bare_Action}
&& S = \frac{1}{2 e_0^2}\,\mbox{Re}\,\mbox{tr}\int d^4x\,
d^2\theta\,W^a W_a + \frac{1}{4} \int d^4x\, d^4\theta\,\phi^{*i}
(e^{\bm{\Omega}^+} e^{2{\cal F}(V)} e^{\bm{\Omega}})_i{}^j \phi_j\nonumber\\
&&\qquad\qquad\qquad\qquad\qquad\qquad\qquad\qquad\qquad\quad +
\Big(\frac{1}{6}\lambda_0^{ijk}\int d^4x\,d^2\theta\, \phi_i
\phi_j \phi_k + \mbox{c.c.}\Big),\qquad
\end{eqnarray}

\noindent
where

\begin{equation}
W_a \equiv \frac{1}{8} \bar D^2 \Big(e^{-\bm{\Omega}} e^{-2{\cal F}(V)} e^{-\bm{\Omega}^+} D_a (e^{\bm{\Omega}^+} e^{2{\cal F}(V)} e^{\bm{\Omega}})\Big).
\end{equation}

\noindent
Under the condition

\begin{equation}\label{Condition_For_Gauge_Invariance}
(T^A)_l{}^k \lambda_0^{ijl}  + (T^A)_l{}^j \lambda_0^{ilk}  + (T^A)_l{}^i \lambda_0^{ljk}  = 0
\end{equation}

\noindent
this action is invariant under the background gauge transformations

\begin{equation}\label{Background_Gauge_Invariance}
e^{\bm{\Omega}} \to e^{iK} e^{\bm{\Omega}} e^{-A};\qquad e^{\bm{\Omega}^+}
\to e^{-A^+} e^{\bm{\Omega}^+} e^{-iK};\qquad V\to e^{iK} V e^{-iK};\qquad \phi \to
e^A \phi,
\end{equation}

\noindent
parameterized by the Lie algebra valued chiral superfield $A$ and the Hermitian superfield $K$. Also it is invariant under the quantum gauge transformations

\begin{equation}\label{Quantum_Invariance}
e^{2{\cal F}(V)} \to e^{-{\cal A}^+} e^{2{\cal F}(V)} e^{-{\cal A}};\qquad
e^{\bm{\Omega}} \to e^{\bm{\Omega}};\qquad e^{\bm{\Omega^+}} \to
e^{\bm{\Omega^+}};\qquad \phi\to e^{-\bm{\Omega}} e^{\cal A}
e^{\bm{\Omega}}\phi.
\end{equation}

\noindent
The quantum gauge invariance is parameterized by the background chiral superfield ${\cal A}$ which satisfies the equation $\bm{\bar \nabla}_{\dot a} {\cal A} = 0$, where

\begin{equation}
\bm{\nabla}_a = e^{-\bm{\Omega}^+} D_a e^{\bm{\Omega}^+}; \qquad \bm{\bar\nabla}_{\dot a} = e^{\bm{\Omega}} \bar D_{\dot a} e^{-\bm{\Omega}}
\end{equation}

\noindent
are the supersymmetric background covariant derivatives.

For regularizing the theory (\ref{Bare_Action}), we add the higher derivative term

\begin{eqnarray}\label{HD_Term}
&& S_{\Lambda} = \frac{1}{2e_0^2}\,\mbox{Re}\,\mbox{tr}\int d^4x\,
d^2\theta\, e^{\bm{\Omega}} W^a e^{-\bm{\Omega}} \Big[R\Big(-\frac{\bar\nabla^2
\nabla^2}{16\Lambda^2}\Big) -1\Big]_{Adj} e^{\bm{\Omega}} W_a e^{-\bm{\Omega}} \qquad\nonumber\\
&& + \frac{1}{4} \int d^4x\, d^4\theta\,\phi^+ e^{\bm{\Omega}^+}
e^{2{\cal F}(V)} \Big[F\Big(-\frac{\bar\nabla^2
\nabla^2}{16\Lambda^2}\Big)-1\Big] e^{\bm{\Omega}} \phi
\end{eqnarray}

\noindent
to its action, where

\begin{equation}
\nabla_a = e^{-2 {\cal F}(V)} \bm{\nabla}_a e^{2{\cal F}(V)}; \qquad \bar\nabla_{\dot a} = \bm{\bar \nabla}_{\dot a}.
\end{equation}

\noindent
The regulator functions $R(x)$ and $F(x)$ rapidly increase at infinity and are equal to 1 at $x=0$; the regularization parameter $\Lambda$ has the dimension of mass and plays the role of an UV cut-off. Also in Eq. (\ref{HD_Term}) we use the notation

\begin{equation}
\Big(a_0 + a_1 V + a_2 V^2 +\ldots\Big)_{Adj} X \equiv a_0 X + a_1 [V,X] + a_2 [V, [V, X]] + \ldots
\end{equation}

In this paper we will use the gauge fixing term invariant under the background gauge transformations\footnote{For simplicity, here in the gauge fixing term we use the same function $R$ as in the higher derivative term $S_\Lambda$.}

\begin{equation}\label{GF_Term}
S_{\mbox{\scriptsize gf}} = -\frac{1}{16 \xi_0 e_0^2}\mbox{tr} \int
d^4x\,d^4\theta\,\bm{\nabla}^2 V R\Big(-\frac{\bm{\bar\nabla}^2
\bm{\nabla}^2}{16\Lambda^2}\Big)_{Adj}\bm{\bar \nabla}^2 V,
\end{equation}

\noindent
where $\xi_0$ is the bare gauge parameter. For constructing the action for the Faddeev--Popov ghosts we define the superfield $\widetilde V = {\cal F}(V)$ (so that $V = {\cal F}^{-1}(\widetilde V)$, where ${\cal F}^{-1}$ is the inverse function) and note that under the infinitesimal quantum gauge transformations

\begin{equation}
\delta V^A = \frac{\partial V^A}{\partial \widetilde V^B}\, \delta \widetilde V^B = \frac{\partial V^A}{\partial \widetilde V^B}\, \left\{\vphantom{\frac{1}{2}}
\smash{\Big(\frac{\widetilde V}{1-e^{2\widetilde V}}\Big)_{Adj} {\cal A}^+ - \Big(\frac{\widetilde V}{1-e^{-2\widetilde V}}\Big)_{Adj}
{\cal A}}\right\}^B,
\end{equation}

\noindent
where $X^A \equiv 2\, \mbox{tr}(X t^A)/e_0$ (or, equivalently, $X=e_0 X^A t^A$). Therefore, the Faddeev--Popov action can be written as

\begin{eqnarray}\label{FP_Action}
&& S_{\mbox{\scriptsize FP}} = \frac{1}{2} \int
d^4x\,d^4\theta\,\frac{\partial {\cal F}^{-1}(\widetilde V)^A}{\partial {\widetilde V}^B}\left.\vphantom{\frac{1}{2}}\right|_{\widetilde V = {\cal F}(V)} \left(e^{\bm{\Omega}}\bar c e^{-\bm{\Omega}} +
e^{-\bm{\Omega}^+}\bar c^+ e^{\bm{\Omega}^+}\right)^A\nonumber\\
&&\qquad\qquad\quad \times \left\{\vphantom{\frac{1}{2}}\smash{
\Big(\frac{{\cal F}(V)}{1-e^{2{\cal F}(V)}}\Big)_{Adj} \Big(e^{-\bm{\Omega}^+} c^+
e^{\bm{\Omega}^+}\Big)  + \Big(\frac{{\cal F}(V)}{1-e^{-2{\cal F}(V)}}\Big)_{Adj}
\Big(e^{\bm{\Omega}} c
e^{-\bm{\Omega}}\Big)}\right\}^B,\qquad
\end{eqnarray}

\noindent
where $c$ and $\bar c$ are the chiral ghost and antighost superfields, respectively. (Therefore, the superfields $c^+$ and $\bar c^+$ are antichiral.)

Also it is necessary to introduce the Nielsen--Kallosh ghosts with the action $S_{\mbox{\scriptsize NK}}$. However, they interact only with the background gauge superfield and are essential only in calculating the one-loop $\beta$-function. That is why here we will not discuss them in detail.

It is important that adding the higher derivative term (\ref{HD_Term}) we do not remove the one-loop divergences. This is a typical feature of the higher derivative regularization, see, e.g., \cite{Faddeev:1980be}. According to \cite{Slavnov:1977zf}, for regularizing divergences in the one-loop approximation it is necessary to insert the Pauli--Villars determinants into the generating functional. In ${\cal N}=1$ supersymmetric gauge theories the one-loop divergences (and subdivergences) coming from loops of the quantum gauge superfield and ghosts can be cancelled by introducing three commuting Pauli--Villars superfields $\varphi_1$, $\varphi_2$, and $\varphi_3$ in the adjoint representation of the gauge group with the action

\begin{eqnarray}
&& S_\varphi = \frac{1}{2e_0^2} \mbox{tr}\int d^4x\, d^4\theta\, \Big(\varphi_1 ^+ \Big[e^{\bm{\Omega}^+} e^{2{\cal F}(V)} R\Big(-\frac{\bar\nabla^2 \nabla^2}{16\Lambda^2}\Big)e^{\bm{\Omega}}\Big]_{Adj}\varphi_1 + \varphi_2^+ \Big[e^{\bm{\Omega}^+} e^{2{\cal F}(V)} e^{\bm{\Omega}}\Big]_{Adj}\varphi_2\qquad\nonumber\\
&& + \varphi_3^+ \Big[e^{\bm{\Omega}^+} e^{2{\cal F}(V)} e^{\bm{\Omega}}\Big]_{Adj}\varphi_3\Big) + \frac{1}{2e_0^2}\Big(\mbox{tr}\int d^4x\, d^2\theta\, M_\varphi (\varphi_1^2 + \varphi_2^2 + \varphi_3^2) +\mbox{c.c.}\Big).\qquad
\end{eqnarray}

\noindent
For this purpose we insert into the generating functional

\begin{equation}
\mbox{Det}(PV,M_\varphi)^{-1} = \int D\varphi_1 D\varphi_2 D\varphi_3 \exp(i S_\varphi).
\end{equation}

\noindent
To cancel the one-loop (sub)divergences coming from the matter loop it is possible \cite{Kazantsev:2017fdc} to use the (commuting) chiral Pauli--Villars superfield in a certain representation $R_{\mbox{\scriptsize PV}}$ for which one can write the gauge invariant mass term, such that $M^{ij} M^*_{jk} = M^2 \delta^i_k$. The corresponding Pauli--Villars determinant which should be inserted into the generating functional can be written as

\begin{equation}
\mbox{Det}(PV,M)^{c} = \Big(\int D\Phi \exp(iS_\Phi)\Big)^{-c},
\end{equation}

\noindent
where $c = T(R)/T(R_{\mbox{\scriptsize PV}})$ and

\begin{equation}
S_\Phi = \frac{1}{4} \int d^4x\, d^4\theta\, \Phi^+ e^{\bm{\Omega}^+} e^{2{\cal F}(V)} F\Big(-\frac{\bar\nabla^2 \nabla^2}{16\Lambda^2}\Big)e^{\bm{\Omega}} \Phi
+ \Big(\frac{1}{4}\int d^4x\, d^2\theta\, M^{ij} \Phi_i \Phi_j +\mbox{c.c.}\Big).
\end{equation}

\noindent
It is important that the masses of the Pauli--Villars superfields $M_\varphi$ and $M$ should be proportional to the parameter $\Lambda$ in the higher derivative term (\ref{HD_Term}),

\begin{equation}
M_\varphi = a_\varphi \Lambda;\qquad M = a\Lambda,
\end{equation}

\noindent
where the constants $a$ and $a_\varphi$ do not depend on the couplings.

Thus, the final expression for the generating functional of the considered theory takes the form

\begin{equation}
Z = \int D\mu\, \mbox{Det}(PV,M_\varphi)^{-1} \mbox{Det}(PV,M)^{c} \exp\Big(i S + i S_\Lambda +i S_{\mbox{\scriptsize gf}} + i S_{\mbox{\scriptsize FP}} + i S_{\mbox{\scriptsize NK}} + i S_{\mbox{\scriptsize sources}}\Big),
\end{equation}

\noindent
where $D\mu$ is the measure of the functional integration and $S_{\mbox{\scriptsize sources}}$ denotes the relevant source terms.

The part of the effective action corresponding to the two-point Green function of the Faddeev--Popov ghosts can be written in the form

\begin{equation}\label{G_C_Definition}
\Gamma^{(2)}_{c} = \frac{1}{4} \int \frac{d^4p}{(2\pi)^4}\, d^4\theta\,\Big(c^{*A}(-p,\theta)\bar c^A(p,\theta)  + \bar c^{*A}(-p,\theta) c^A(p,\theta) \Big)
G_c(\alpha_0,\lambda_0,\xi_0,y_0,\ldots,\Lambda/p),
\end{equation}

\noindent
where dots denote the other parameters describing the nonlinear renormalization.\footnote{Below in all equations, for simplicity, we will omit these dots and explicitly write only dependence of the renormalization group functions on $y$, which is essential in the considered approximation.} In our notation the renormalization constants are defined by the equations\footnote{Here we present only the renormalization constants needed in this paper. Some other definitions can be found in \cite{Aleshin:2016yvj}.}

\begin{equation}
\frac{1}{\alpha_0} = \frac{Z_\alpha}{\alpha};\qquad
\frac{1}{\xi_0} = \frac{Z_\xi}{\xi};\qquad  \bar c c = Z_c Z_\alpha^{-1} \bar c_R c_R;\qquad V = Z_V Z_\alpha^{-1/2} V_R;\qquad y_0 = Z_y y,
\end{equation}

\noindent
where $\alpha_0 \equiv e_0^2/4\pi$ and the subscript $R$ denotes the renormalized superfields. This implies that if $V=e_0V^A t^A$ and $V_R = e V_R^A t^A$, then $V^A = Z_V V^A_R$. Moreover, in this notation the function $Z_c G_c$ (expressed in terms of the renormalized couplings) is finite in the limit $\Lambda\to \infty$.

The one-loop renormalization of the considered theory has been investigated in Ref. \cite{Aleshin:2016yvj}. In particular, (for $y_0=0$) the one-loop anomalous dimension of the Faddeev--Popov ghost superfields defined in terms of the bare couplings is

\begin{equation}\label{Gamma_C}
\gamma_c(\alpha_0,\lambda_0,\xi_0,y_0=0) \equiv \left. -\frac{d\ln Z_c}{d\ln\Lambda}\right|_{\alpha,\lambda,\xi,y=\mbox{\scriptsize const}} =  \frac{\alpha_0 C_2 (\xi_0-1)}{6\pi} + O(\alpha_0^2,\alpha_0\lambda_0^2).
\end{equation}

\noindent
Also below we will need the one-loop running of the gauge coupling constant and of the gauge parameter which is encoded in the renormalization group functions

\begin{eqnarray}\label{Beta_Function}
&& \beta(\alpha_0,\lambda_0,\xi_0,y_0=0)\equiv \left.\frac{d\alpha_0}{d\ln\Lambda}\right|_{\alpha,\lambda,\xi,y=\mbox{\scriptsize const}} = -\frac{\alpha_0^2}{2\pi}\Big(3C_2 - T(R) + O(\alpha_0,\lambda_0^2)\Big);\qquad\\
\label{Xi_Renormalization}
&& \gamma_\xi(\alpha_0,\lambda_0,\xi_0,y_0=0) \equiv \left.- \frac{d\ln Z_\xi}{d\ln\Lambda} \right|_{\alpha,\lambda,\xi,y=\mbox{\scriptsize const}} = 2 \left.\frac{d\ln Z_V}{d\ln\Lambda} \right|_{\alpha,\lambda,\xi,y=\mbox{\scriptsize const}}\qquad\nonumber\\
&& = -2 \gamma_V(\alpha_0,\lambda_0,\xi_0,y_0=0) = \frac{\alpha_0 C_2 (\xi_0-1)}{3\pi} - \frac{\beta(\alpha_0,\lambda_0,\xi_0,y_0=0)}{\alpha_0}
+ O(\alpha_0^2,\alpha_0\lambda_0^2),\qquad
\end{eqnarray}

\noindent
respectively. Expressing the $\beta$-function in terms of the renormalization constant $Z_\alpha$, from these equations we obtain (again for $y_0=0$)

\begin{equation}
\left.\frac{d\ln (Z_\alpha Z_\xi)}{d\ln\Lambda}\right|_{\alpha,\lambda,\xi,y=\mbox{\scriptsize const}} = -\frac{\alpha_0 C_2 (\xi_0-1)}{3\pi} + O(\alpha_0^2,\alpha_0\lambda_0^2).
\end{equation}

\noindent
However, as we will see below, for calculating the two-loop anomalous dimension of the Faddeev--Popov ghosts it is necessary to take into account the non-linear renormalization. In the lowest-order approximation it corresponds to the one-loop renormalization of the parameter $y$, which can be found from the results of Refs. \cite{Juer:1982fb,Juer:1982mp}. Note that here we use the higher covariant derivative regularization, while in Refs. \cite{Juer:1982fb,Juer:1982mp} the theory was regularized by dimensional reduction. However, in the one-loop approximation for an arbitrary theory $1/\varepsilon$ (where $\varepsilon\equiv 4-D$) within the dimensional technique corresponds to $\ln\Lambda$ in the case of using the higher covariant derivative regularization, see, e.g., \cite{Chetyrkin:1980sa,Pronin:1997eb}. That is why the result of \cite{Juer:1982fb,Juer:1982mp} can be presented in the form

\begin{equation}\label{Y_Renormalization}
y_0 = y + \frac{\alpha}{90\pi} \Big((2+3\xi) \ln\frac{\Lambda}{\mu} + k_1\Big) + \ldots,
\end{equation}

\noindent
where $k_1$ is an arbitrary finite constant. Dots denote the one-loop terms containing coefficients of the nonlinear part of the function ${\cal F}(V)$ and the higher order contributions. Note that writing Eq. (\ref{Y_Renormalization}) we take into account the difference between our notations and the notations of Refs. \cite{Juer:1982fb,Juer:1982mp},

\begin{equation}
\xi \sim \alpha;\quad\ \ t^A \sim \frac{1}{\sqrt{2}} G_i;\quad\ \ f^{ABC} \sim \frac{1}{\sqrt{2}} f_{ijk};\quad\ \ G^{ABCD} \sim \frac{1}{4} G_{ijkl};\quad\ \ V^A \sim \frac{1}{\sqrt{2}} V^i.
\end{equation}

\noindent
Eq. (\ref{Y_Renormalization}) can be equivalently presented in the form

\begin{equation}
\gamma_y(\alpha_0,\lambda_0,\xi_0,y_0) = - \left.\frac{d\ln Z_y}{d\ln \Lambda}\right|_{\alpha,\lambda,\xi,y=\mbox{\scriptsize const}} = - \frac{\alpha_0 (2+3\xi_0)}{90\pi y_0} + \ldots
\end{equation}

\section{Two-loop ghost anomalous dimension defined in terms of the bare couplings}
\hspace*{\parindent}\label{Section_Anomalous_Dimension}

In this section we will calculate the two-loop anomalous dimension of the Faddeev--Popov ghosts defined in terms of the bare couplings. It can be written as a derivative of the logarithm of the function $G_c$ defined by Eq. (\ref{G_C_Definition}) with respect to $\ln\Lambda$ in the limit of the vanishing external momentum,

\begin{equation}\label{Gamma_Ghost_Definition}
\gamma_c(\alpha_0,\lambda_0,\xi_0,y_0) \equiv \left. -\frac{d\ln Z_c}{d\ln\Lambda}\right|_{\alpha,\lambda,\xi,y=\mbox{\scriptsize const}} = \left. \frac{d\ln G_c}{d\ln\Lambda}\right|_{\alpha,\lambda,\xi,y=\mbox{\scriptsize const};\, p=0}.
\end{equation}

\noindent
The last equality follows from the finiteness of the function $Z_c G_c$ expressed in terms of the renormalized quantities, while the condition $p\to 0$ is needed for removing finite terms proportional to $(p/\Lambda)^k$, where $k$ is a positive integer.

\begin{figure}[h]
\begin{picture}(0,7)
\put(0,5.4){\includegraphics[scale=0.14]{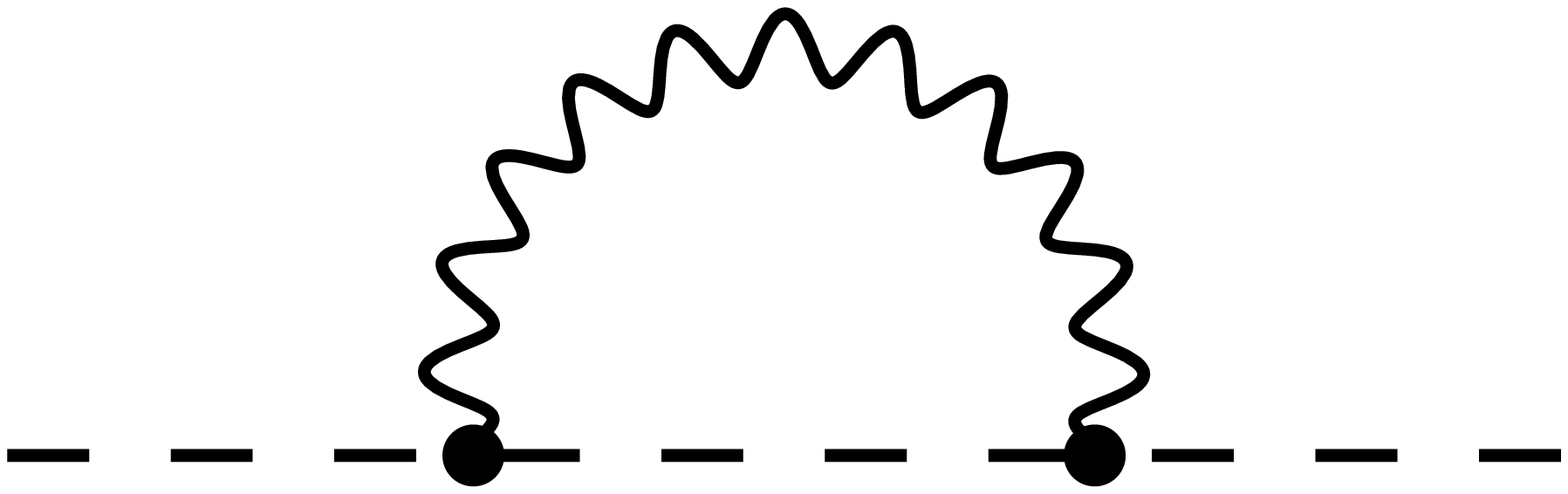}}
\put(0,6.5){$(1)$}
\put(3.3,5.4){\includegraphics[scale=0.14]{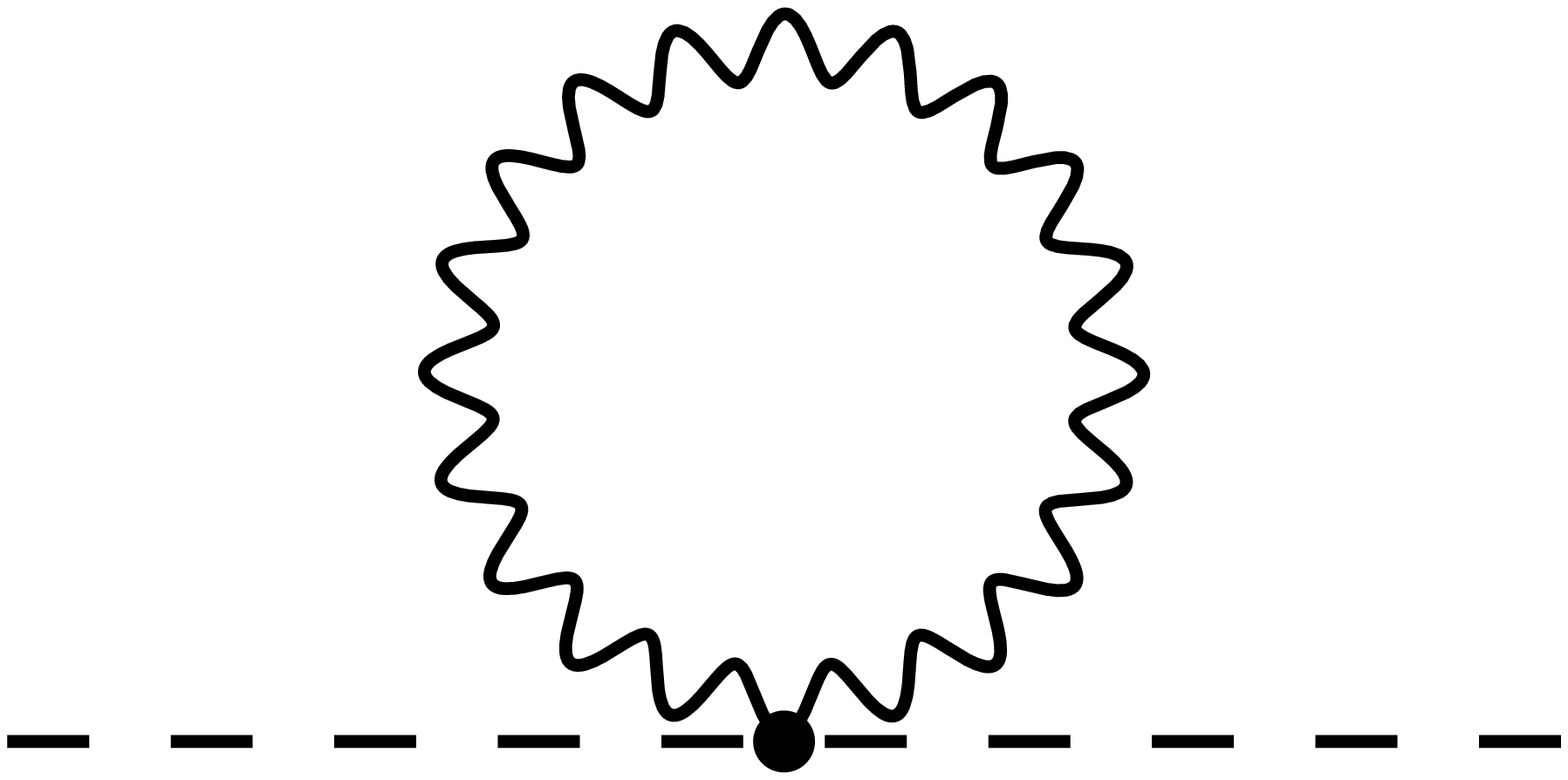}}
\put(3.3,6.5){$(2)$}
\put(6.6,5.4){\includegraphics[scale=0.14]{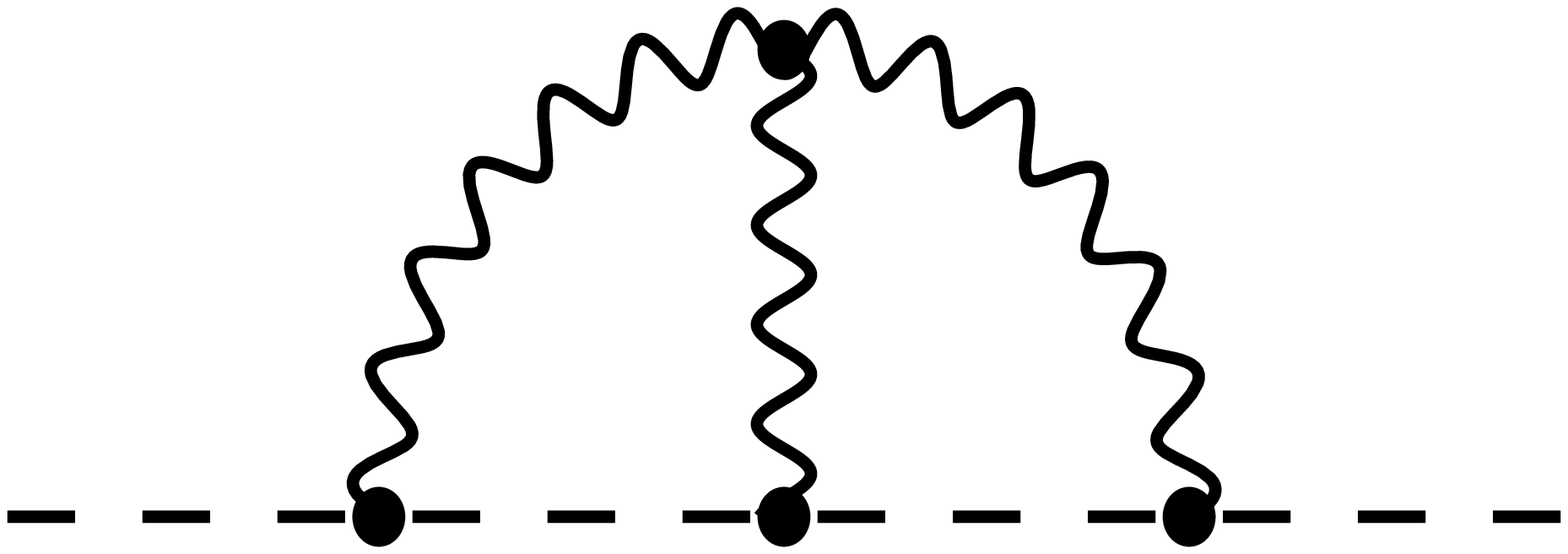}}
\put(6.6,6.5){$(3)$}
\put(9.9,5.4){\includegraphics[scale=0.14]{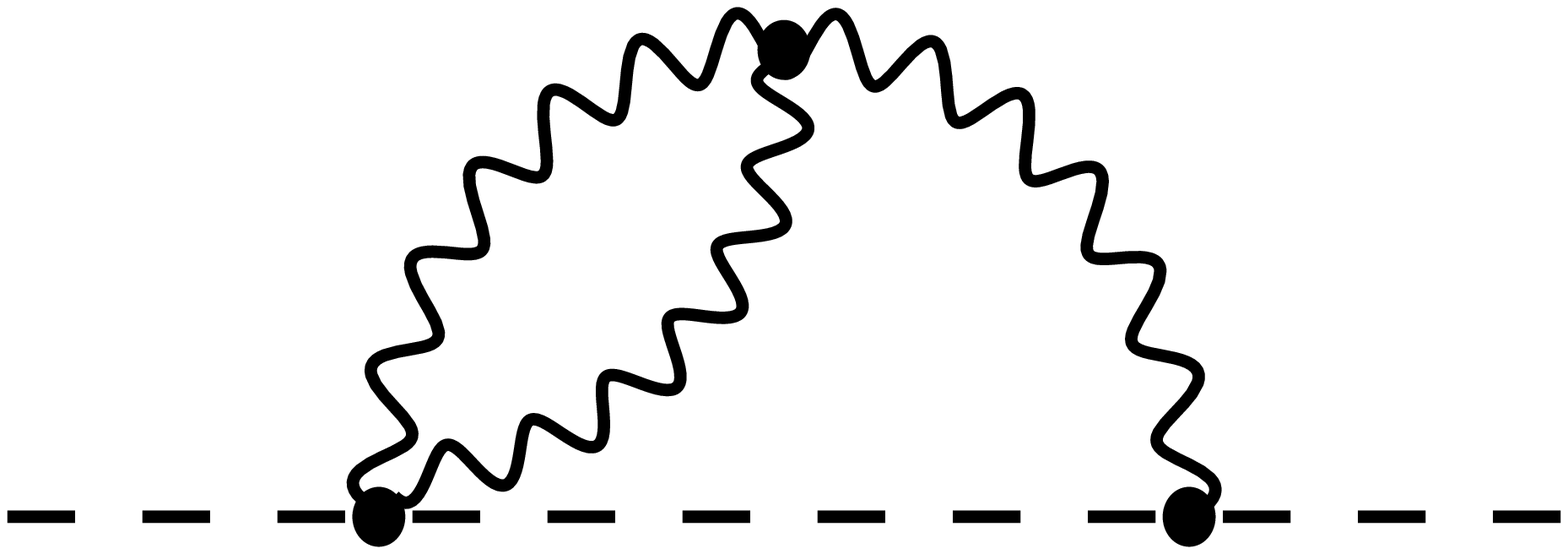}}
\put(9.9,6.5){$(4)$}
\put(13.2,5.4){\includegraphics[scale=0.14]{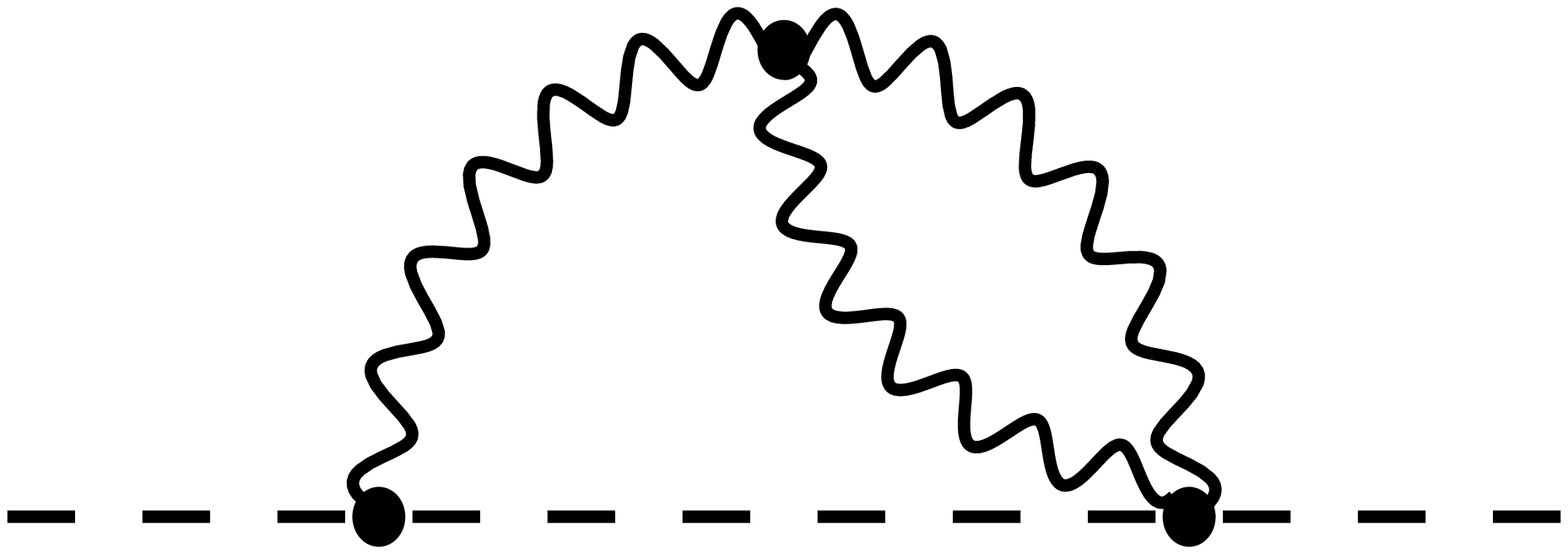}}
\put(13.2,6.5){$(5)$}
\put(0,2.7){\includegraphics[scale=0.14]{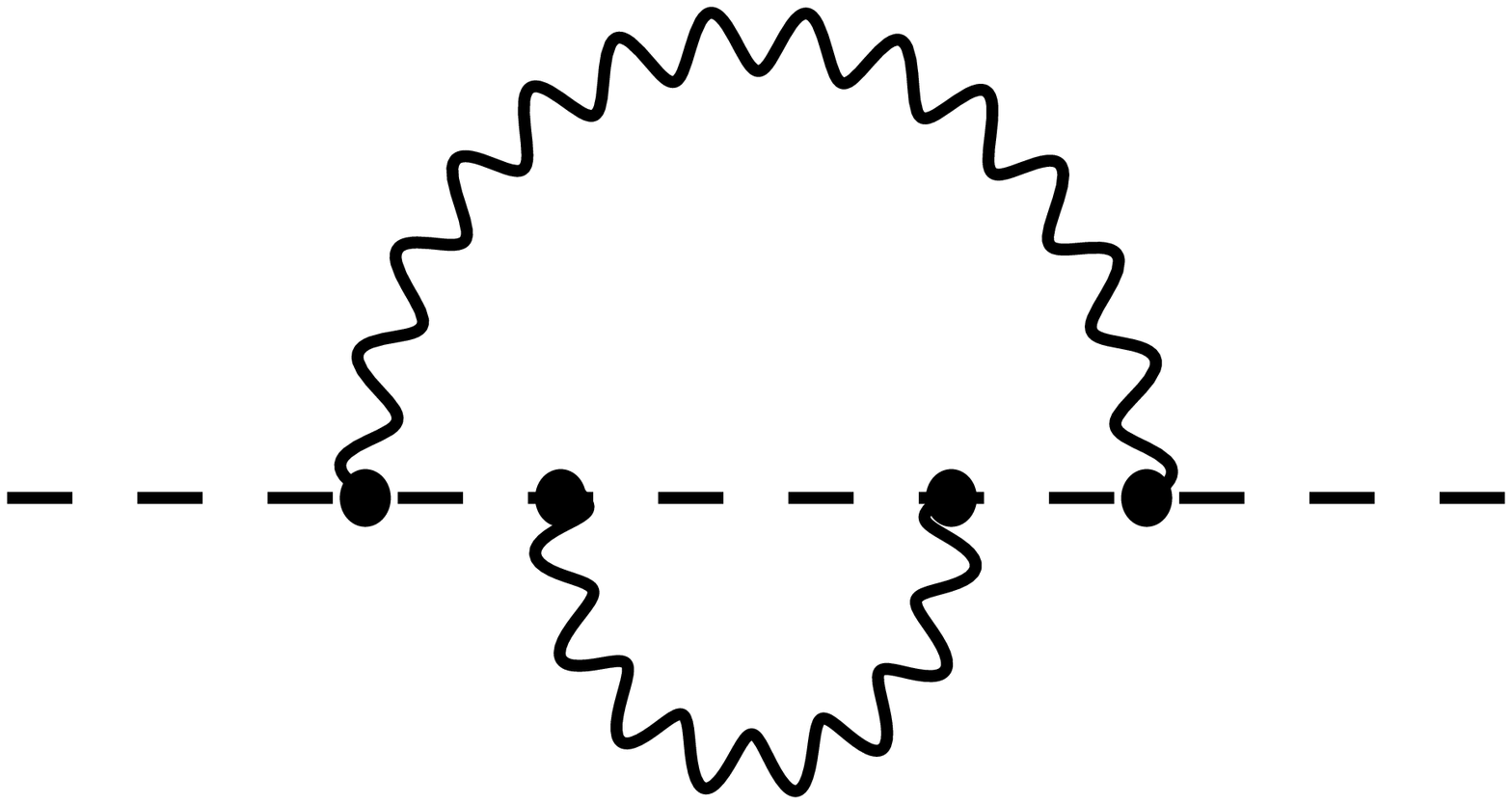}}
\put(0,4.0){$(6)$}
\put(3.3,2.6){\includegraphics[scale=0.14]{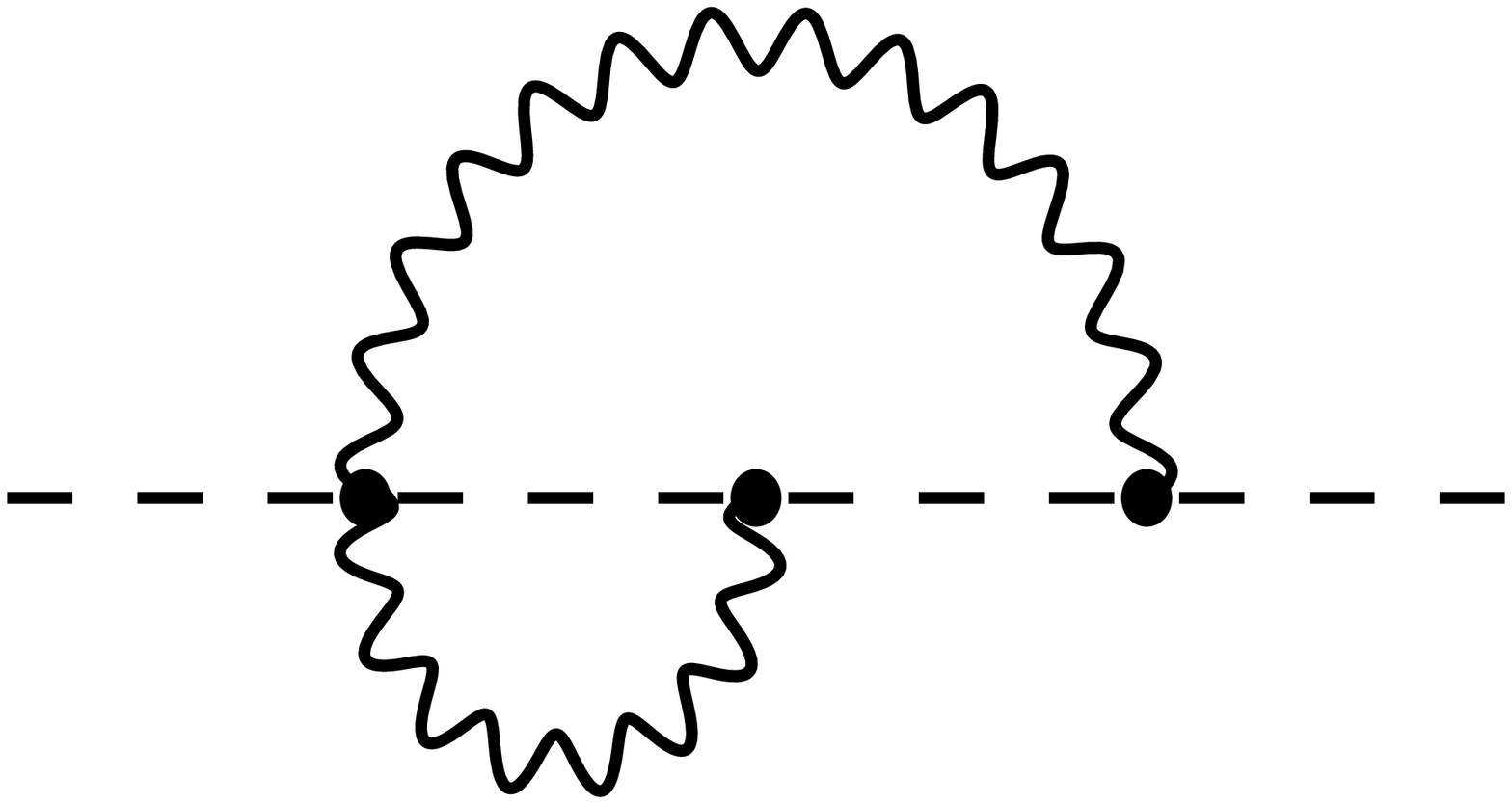}}
\put(3.3,4.0){$(7)$}
\put(6.6,2.6){\includegraphics[scale=0.14]{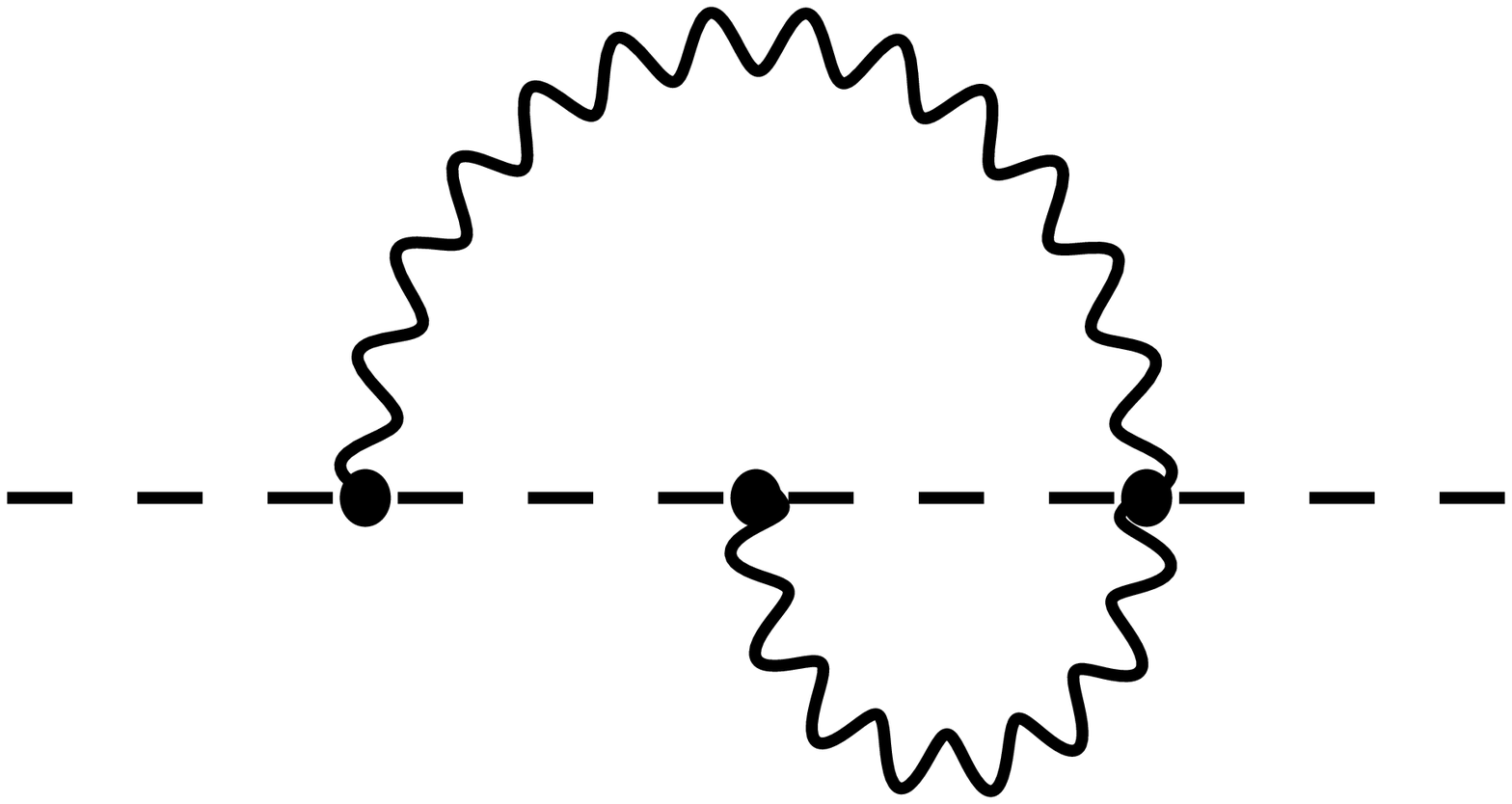}}
\put(6.6,4.0){$(8)$}
\put(9.9,2.6){\includegraphics[scale=0.14]{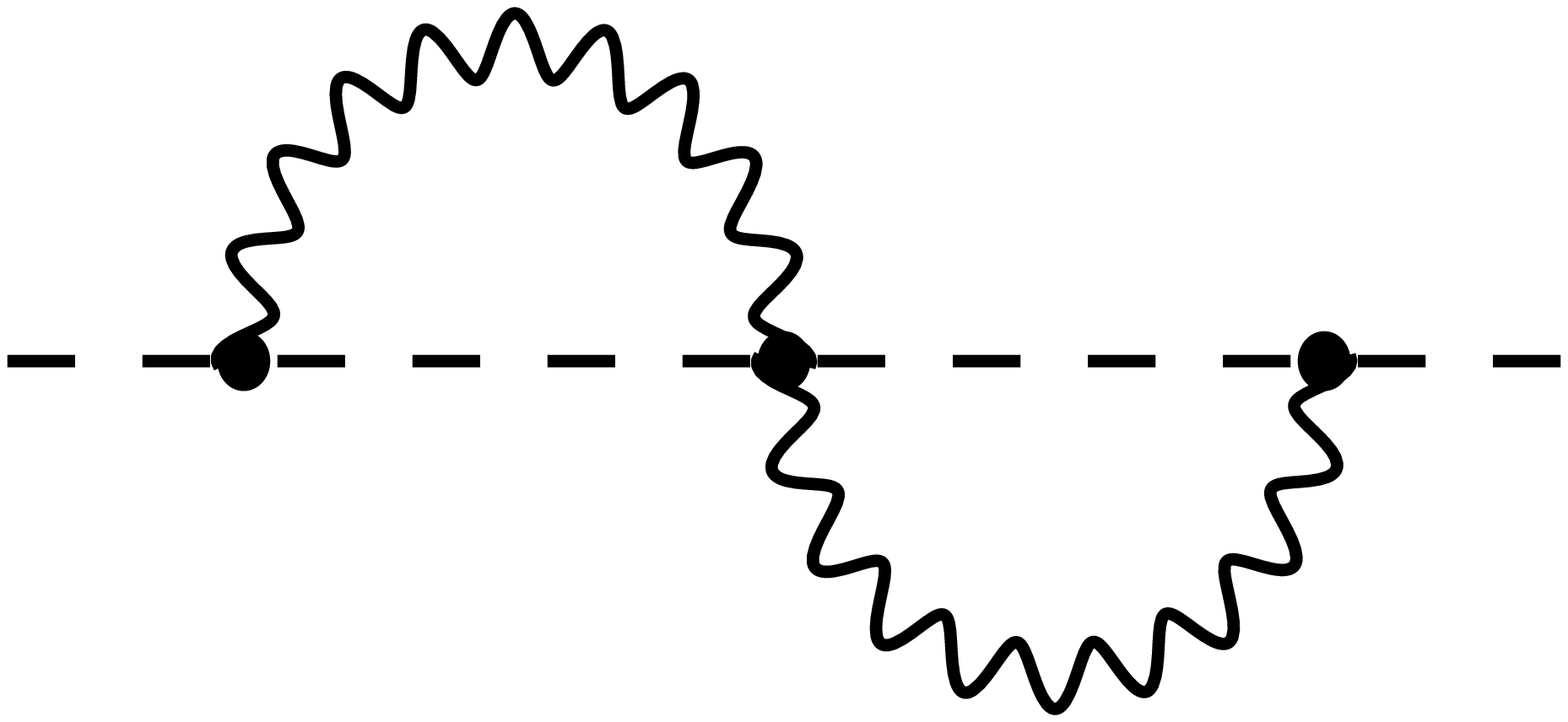}}
\put(9.9,4.0){$(9)$}
\put(13.2,2.5){\includegraphics[scale=0.14]{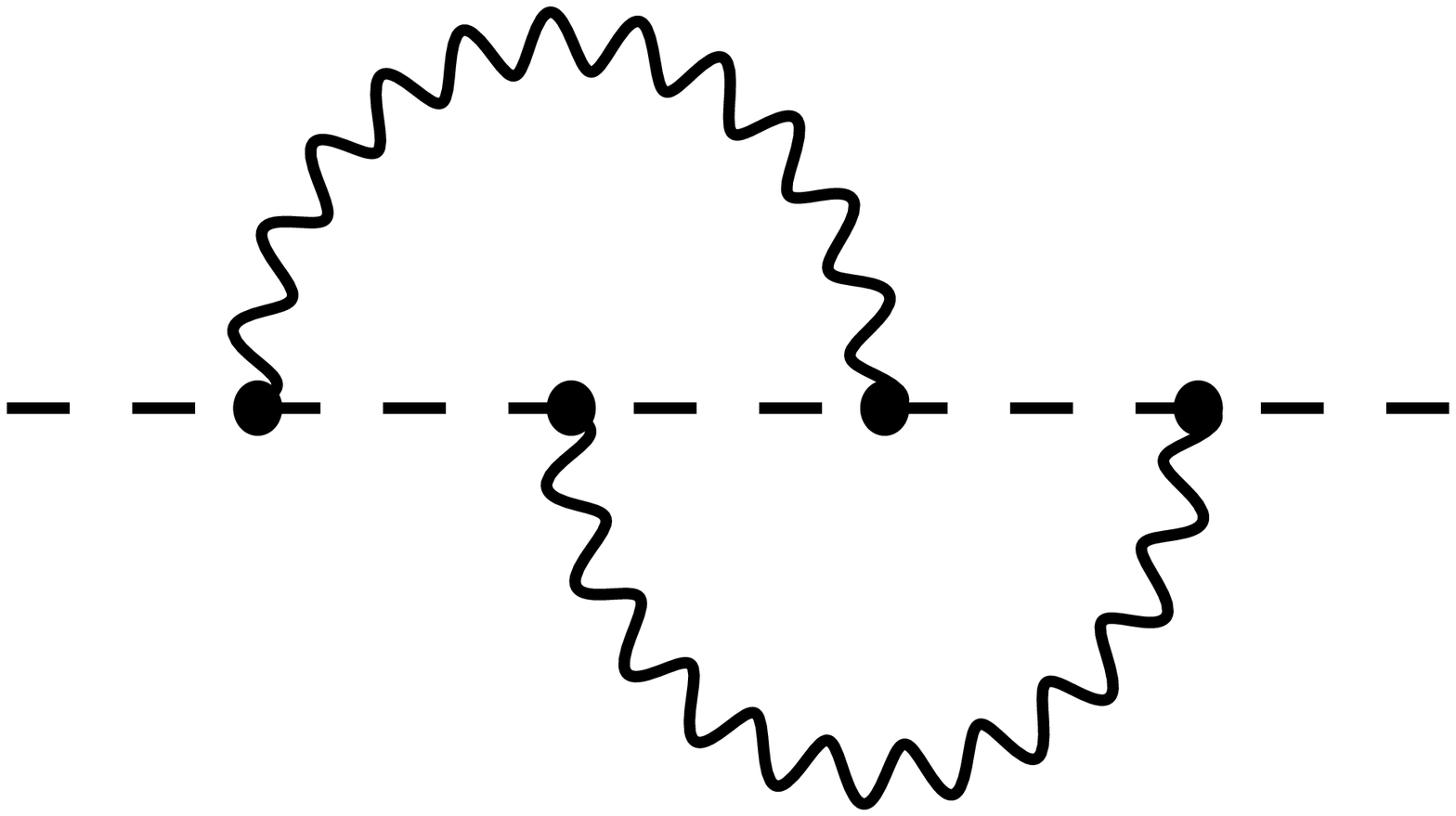}}
\put(13.2,4.0){$(10)$}
\put(0,0){\includegraphics[scale=0.14]{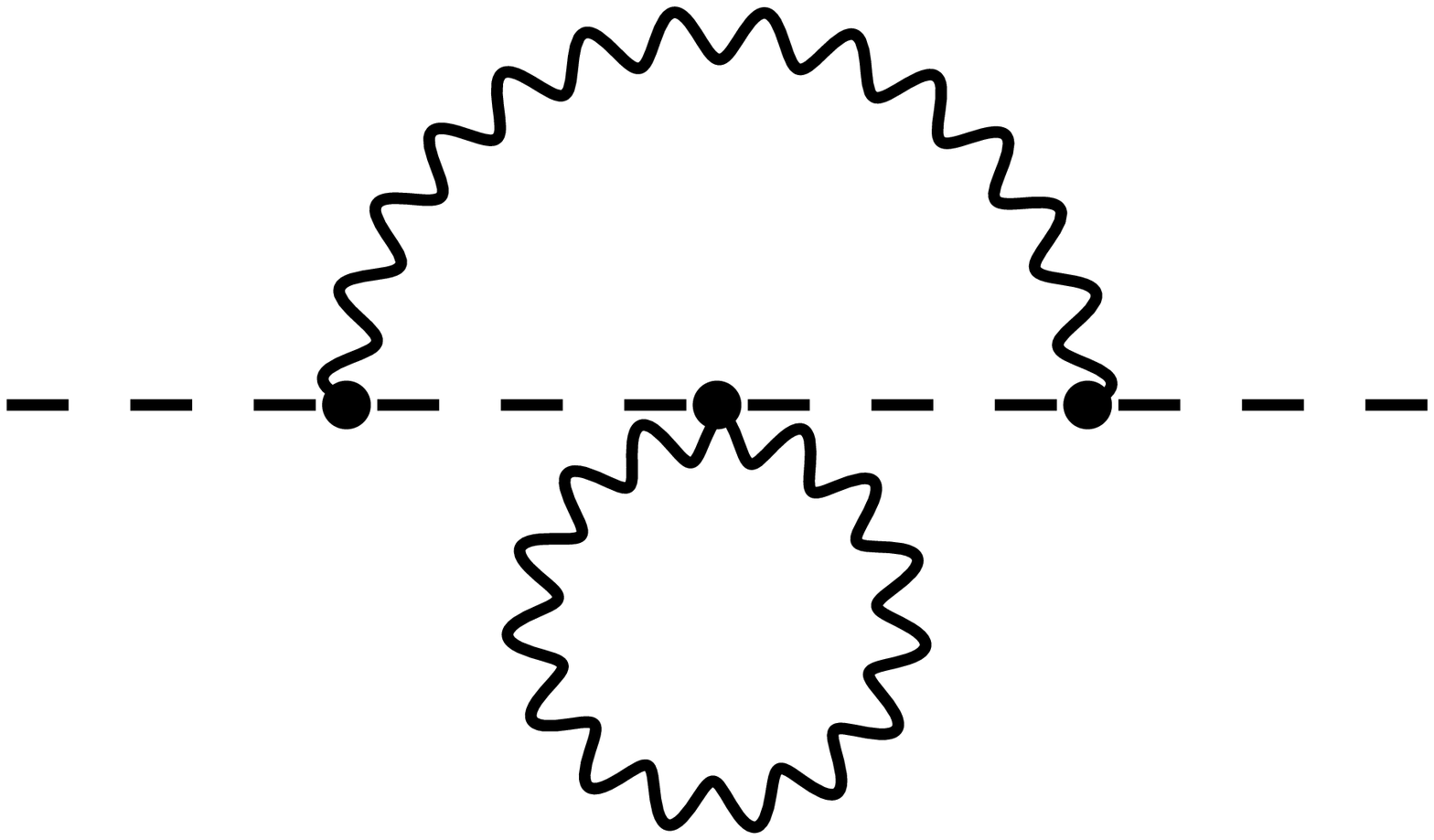}}
\put(0,1.6){$(11)$}
\put(3.3,0.17){\includegraphics[scale=0.14]{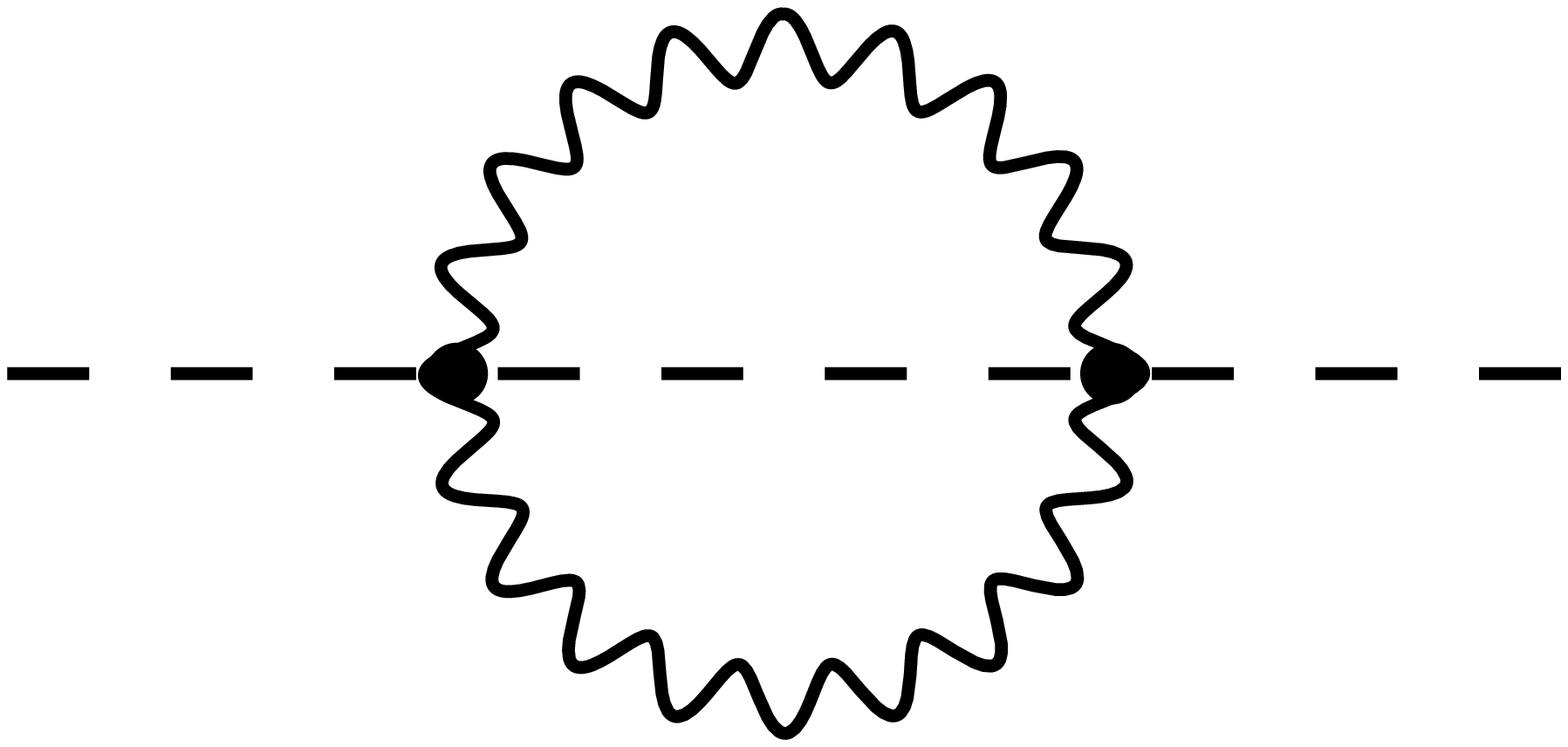}}
\put(3.3,1.6){$(12)$}
\put(6.6,-0.3){\includegraphics[scale=0.14]{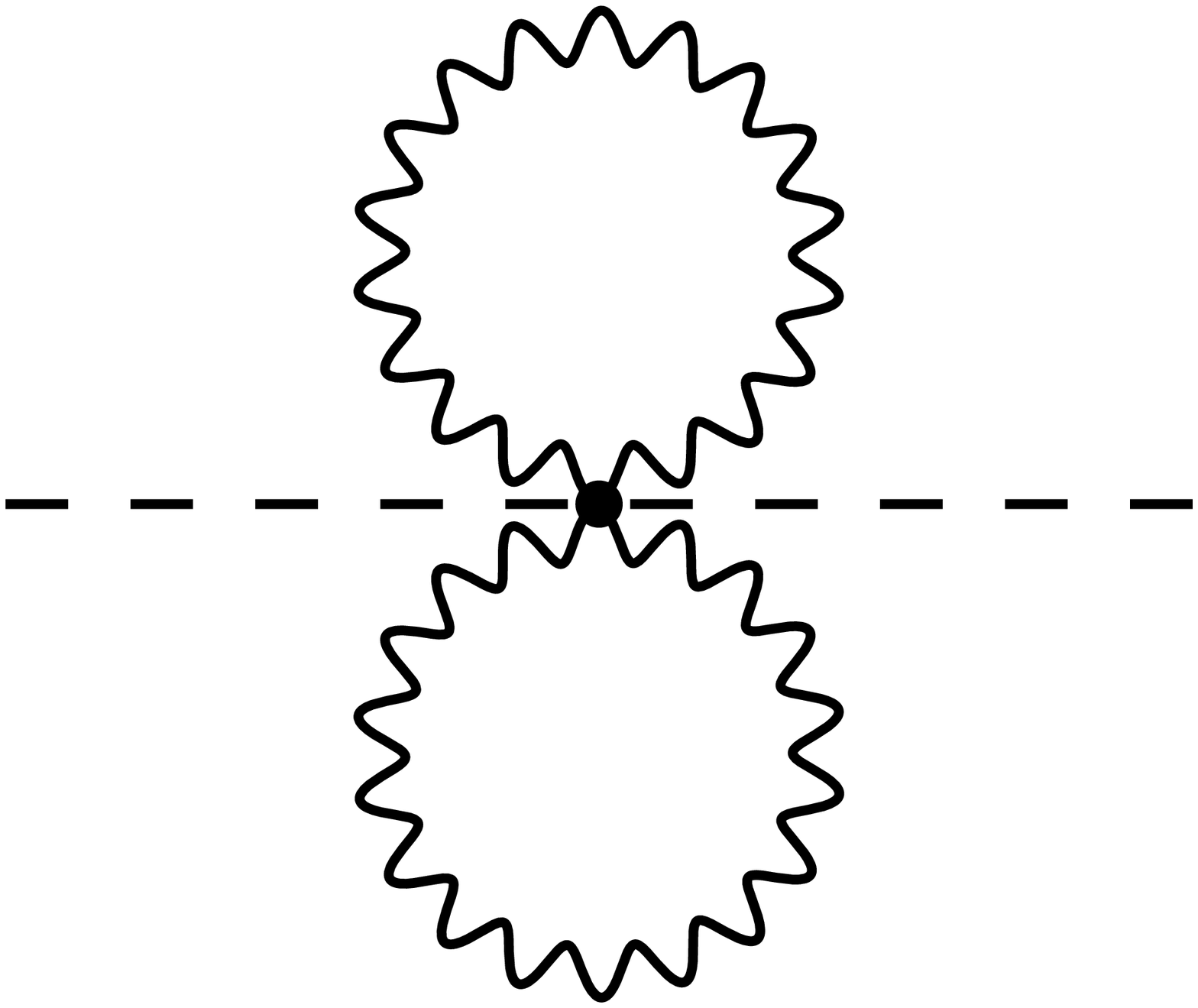}}
\put(6.6,1.6){$(13)$}
\put(9.9,0.1){\includegraphics[scale=0.14]{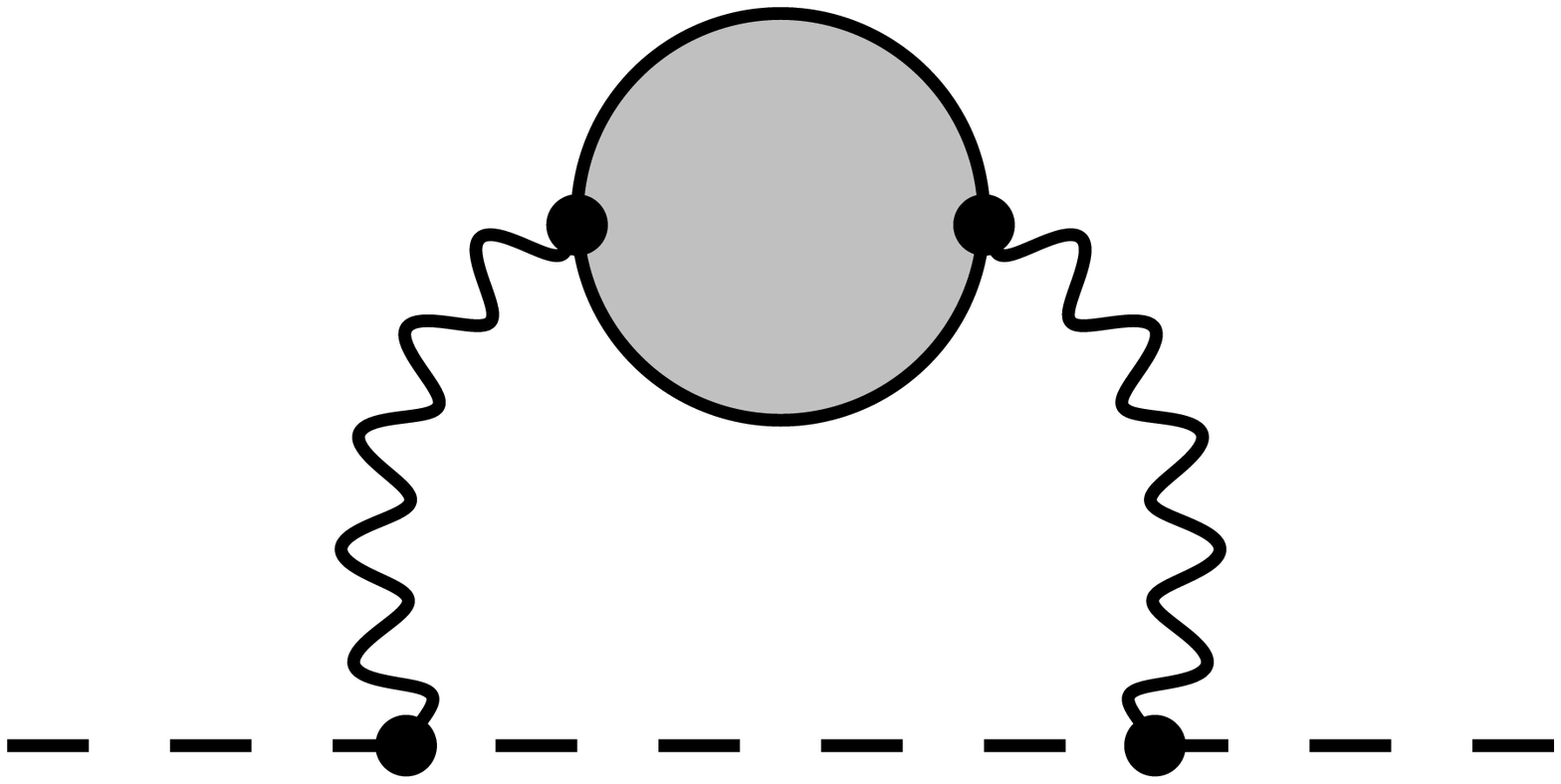}}
\put(9.9,1.6){$(14)$}
\put(13.2,0.1){\includegraphics[scale=0.14]{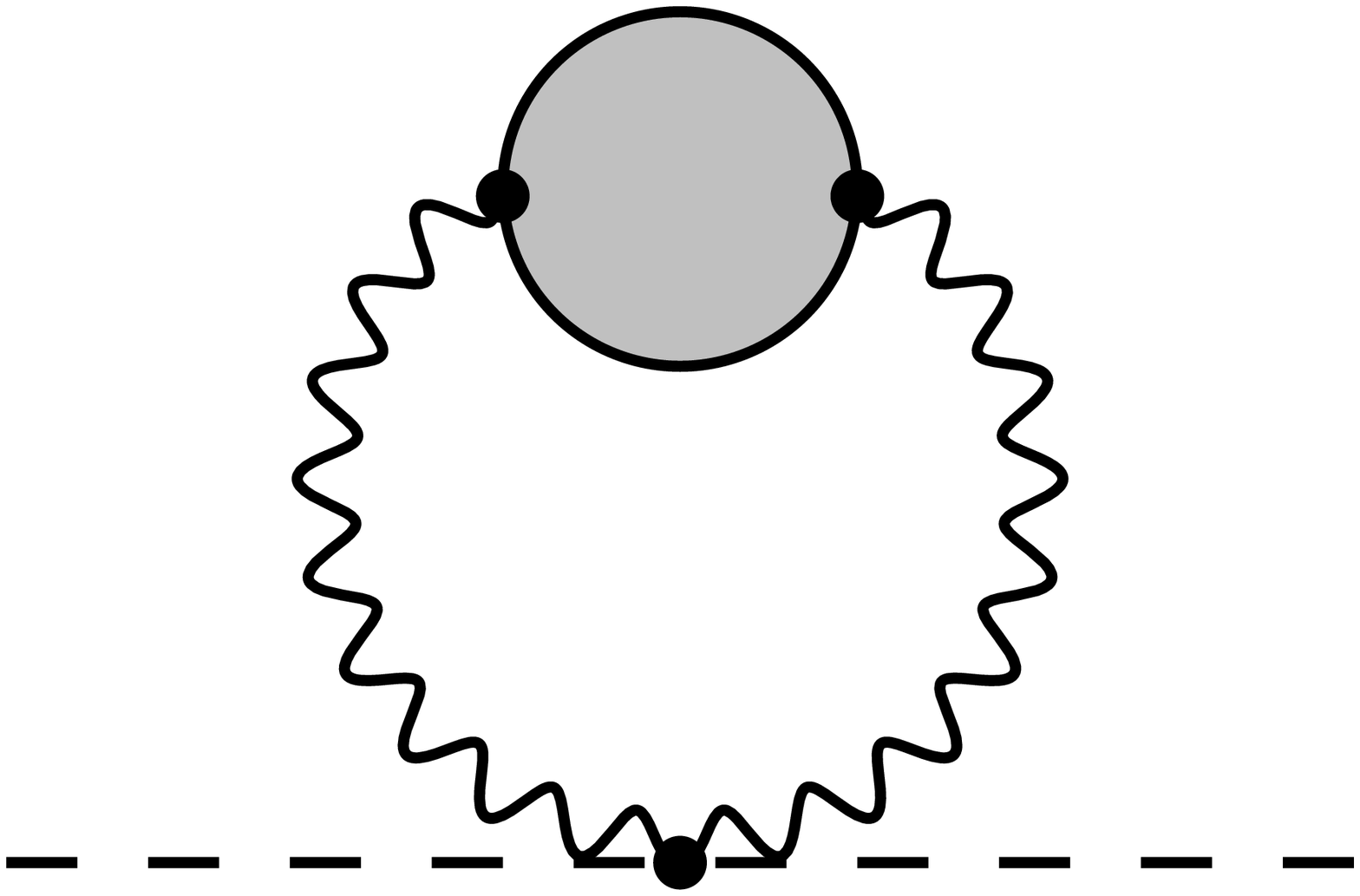}}
\put(13.2,1.6){$(15)$}
\end{picture}
\caption{One- and two-loop superdiagrams contributing to the two-point Green function of the Faddeev--Popov ghosts.}\label{Figure_Diagrams}
\end{figure}

It is important that we will calculate $\gamma_c(\alpha_0,\lambda_0,\xi_0,y_0)$ for the case ${\cal F}(V) = V$. This implies that $y_0$ and similar constants corresponding to the higher powers of $V$ in Eq. (\ref{Nonlinear_Function}) are set to 0 in the final result. Actually, this corresponds to a special choice of the gauge condition, because after the substitution $\widetilde V = {\cal F}(V)$ the parameters of the nonlinear renormalization become gauge parameters. Equivalence of a nonlinear renormalization to a nonlinear change of the gauge was demonstrated in \cite{Piguet:1981fb}. However, as we will see below, the presence of the parameter $y_0$ at intermediate steps of the calculation is very important. The matter is that it is necessary to take into account the one-loop renormalization of this parameter, while in the two-loop diagrams in the case ${\cal F}(V) = V$ with the considered accuracy it can be set to 0 together with the other similar parameters. Having this in mind, we can construct the superdiagrams contributing to the two-point Green function of the Faddeev--Popov ghosts in the considered approximation. They are presented in Fig. \ref{Figure_Diagrams}. In this figure dashed lines denote ghost propagators and external lines, while wavy lines denote propagators of the gauge superfield $V$. We assume that in all diagrams the left ghost leg corresponds to the antighost $\bar c^+$ and the right ghost leg corresponds to the ghost $c$. In the diagrams (14) and (15) the gray circles denote insertions of the one-loop polarization operator, which has been calculated in \cite{Kazantsev:2017fdc}.

The ghost vertices entering the diagrams in Fig. \ref{Figure_Diagrams} can be constructed from Eq. (\ref{FP_Action}) using the equations

\begin{eqnarray}
&& \frac{V}{1-e^{2V}} = -\frac{1}{2} + \frac{1}{2} V - \frac{1}{6} V^2 + \frac{1}{90} V^4 + O(V^6);\\
&& \frac{V}{1-e^{-2V}} = \frac{1}{2} + \frac{1}{2} V + \frac{1}{6} V^2 - \frac{1}{90} V^4 + O(V^6);\\
&& {\cal F}^{-1}(\widetilde V)^A = \widetilde V^A - e_0^2\, y_0\, G^{ABCD} \widetilde V^B \widetilde V^C \widetilde V^D + \ldots \vphantom{\frac{1}{2}}
\end{eqnarray}

\noindent
The relevant terms of Eq. (\ref{FP_Action}) can be written as

\begin{eqnarray}\label{FP_Action_Lower_Terms}
&& S_{\mbox{\scriptsize FP}} = \int d^4x\,d^4\theta\, \left(\frac{1}{4} c^{*A} \bar c^A + \frac{1}{4} \bar c^{*A} c^A
+ \frac{ie_0}{4} f^{ABC} (\bar c^A + \bar c^{*A}) V^B (c^C + c^{*C})  - \frac{e_0^2}{12} f^{ABC}\right.\nonumber\\
&& \times f^{CDE} (\bar c^A + \bar c^{*A}) V^B V^D (c^E - c^{*E})
- \frac{e_0^4}{180} f^{ABC} f^{CDE} f^{EFG} f^{GHI} (\bar c^A + \bar c^{*A}) V^B V^D V^F \qquad
\nonumber\\
&&\left. \times V^H (c^I - c^{*I}) -\frac{3}{4} e_0^2\, y_0\, G^{ABCD} (\bar c^A + \bar c^{*A}) V^C V^D (c^B-c^{*B}) + \ldots
\right).
\end{eqnarray}

\noindent
Note that the quintic vertices containing two external ghost legs and three legs of the quantum gauge superfield (which may arise due to the nonlinear form of the function ${\cal F}(V)$) are not essential in the considered approximation. In principle, they can appear in the two-loop diagrams, but the corresponding contributions to the two-loop ghost anomalous dimension vanish if we choose ${\cal F}(V) = V$.

Expressions for all diagrams presented in Fig. \ref{Figure_Diagrams} are listed in Appendix \ref{Appendix_Superdiagrams_For_Gc}. Note that, as we discussed above, calculating these diagrams we take into account the dependence on $y_0$ only in the one-loop approximation. This is necessary, because the renormalization of the parameter $y_0$ is important for calculating the anomalous dimension. In the two-loop approximation $y_0$-dependence can be ignored due to the condition ${\cal F}(V) = V$ which will be always assumed.

Most of the superdiagrams presented in Fig. \ref{Figure_Diagrams} vanish in the limit $p\to 0$. Nontrivial results have been obtained only for the diagrams (2), (8), (12), (13), and (15),

\begin{eqnarray}\label{Gamma_C_General}
&&\hspace*{-4mm} \gamma_c(\alpha_0,\lambda_0,\xi_0,y_0) = \frac{d\ln G_c}{d\ln\Lambda}\left.\vphantom{\frac{1}{2}}\right|_{\alpha,\lambda,\xi,y=\mbox{\scriptsize const};\, p\to 0} = \frac{d}{d\ln\Lambda}\left(
\Delta G_c^{(2)} -\frac{1}{2}\left(\Delta G_c^{(2)}\right)^2 + \Delta G_c^{(8)} \right.\nonumber\\
&&\hspace*{-4mm}\left. + \Delta G_c^{(12)} + \Delta G_c^{(13)} + \Delta G_c^{(15)}\right)\Big|_{\alpha,\lambda,\xi,y=\mbox{\scriptsize const}} + \dots,
\end{eqnarray}

\noindent
where dots denote two-loop terms containing parameters describing nonlinear renormalization of the quantum gauge superfield and the higher order terms. $\Delta G_c^{(A)}$ denotes the contribution of the diagram $A$ in Fig. \ref{Figure_Diagrams} to the function $G_c$ in the limit of the vanishing external momentum. Due to this limit the loop integrals inside $\Delta G_c^{(A)}$ are not well defined. However, the expression in the right hand side of Eq. (\ref{Gamma_C_General}) is well defined due to the differentiation with respect to $\ln\Lambda$ (which should be done before integrations). Substituting explicit expressions for $\Delta G_c^{(A)}$ presented in Appendix \ref{Appendix_Superdiagrams_For_Gc}, after some transformations, we obtain\footnote{We omit terms proportional to $\alpha_0^2\, y_0$ due to the condition ${\cal F}(V)=V$.}

\begin{eqnarray}\label{Gamma_C_Integral}
&&\hspace*{-5mm} \gamma_c = \frac{d}{d\ln\Lambda}\Bigg\{ 4\pi C_2  \int \frac{d^4k}{(2\pi)^4} \frac{\alpha_0}{k^4 R_k}\left[(\xi_0-1) \Big(\frac{1}{3} - \frac{5}{2} y_0 C_2\Big) + \frac{8\pi\alpha_0}{3 R_k}\Big(C_2 f(k/\Lambda)+ T(R) h(k/\Lambda)\Big) \right] \nonumber\\
&&\hspace*{-5mm} + 4\pi^2 C_2^2 \alpha_0^2 \int \frac{d^4k}{(2\pi)^4} \frac{d^4l}{(2\pi)^4} \frac{1}{R_k R_l} \left[\frac{(\xi_0-1)(5\xi_0+8)}{9 k^4 l^4}  - \frac{4(\xi_0^2-1)}{3 k^4 l^2 (k+l)^2} \right] \Bigg\}\Bigg|_{\alpha,\xi,y=\mbox{\scriptsize const}}+\ldots,
\end{eqnarray}

\noindent
where $R_k \equiv R(k^2/\Lambda^2)$, and the functions $f(k/\Lambda)$ and $h(k/\Lambda)$ are related to the one-loop polarization operator of the quantum gauge superfield $V$ calculated in \cite{Kazantsev:2017fdc}. For completeness, we present explicit expressions for these functions in Appendix \ref{Appendix_Loop_Integrals}. It is important that the expression (\ref{Gamma_C_Integral}) should be evaluated at fixed values of the renormalized couplings $\alpha$, $\xi$, and $y$. This implies that we should express the bare couplings in terms of the renormalized ones by the help of the equations

\begin{eqnarray}\label{Alpha_Renormalization}
&& \alpha_0 = \alpha - \frac{\alpha^2}{2\pi}\Big[3 C_2 \Big(\ln\frac{\Lambda}{\mu} + b_{11}\Big) - T(R)\Big(\ln\frac{\Lambda}{\mu}+ b_{12}\Big)\Big] + O(\alpha^3,\alpha^2\lambda^2);\qquad\\
\label{Alpha_Xi_Renormalization}
&& \alpha_0 \xi_0 = \alpha\xi + \frac{\alpha^2 C_2}{3\pi} \Big(\xi(\xi-1)\ln\frac{\Lambda}{\mu} + x_1\Big) + O(\alpha^3, \alpha^2\lambda^2);\\
\label{Y_Renormalization_Copy}
&& y_0 = y + \frac{\alpha}{90\pi} \Big((2+3\xi) \ln\frac{\Lambda}{\mu} + k_1\Big) + \ldots,
\end{eqnarray}

\noindent
where the finite constants $b_{11}$, $b_{12}$, $x_1$, and $k_1$ appear due to arbitrariness of choosing a subtraction scheme. Substituting the bare couplings from Eqs. (\ref{Alpha_Renormalization}), (\ref{Alpha_Xi_Renormalization}), and (\ref{Y_Renormalization_Copy}) and calculating the remaining integrals we find the anomalous dimension $\gamma_c$ in the considered approximation. The details of this calculation are presented in Appendix \ref{Appendix_Calculation}. The result has been obtained for the higher derivative regulator function

\begin{equation}
R(x) = 1+x^n
\end{equation}

\noindent
(where $n\ge 1$ is a positive integer) and can be written as

\begin{eqnarray}\label{Bare_Gamma_With_Y}
&&\hspace*{-5mm} \gamma_c(\alpha_0,\lambda_0,\xi_0,y_0) = \frac{\alpha_0 C_2 (\xi_0-1)}{6\pi} - \frac{5\alpha_0 y_0 C_2^2 (\xi_0-1)}{4\pi}  -  \frac{\alpha_0^2 C_2^2}{24\pi^2} \Big(\xi_0^2-1\Big) - \frac{\alpha_0^2 C_2^2}{4\pi^2}\Big( \ln a_\varphi + 1\Big) \nonumber\\
&&\hspace*{-5mm} + \frac{\alpha_0^2 C_2 T(R)}{12\pi^2}\Big( \ln a + 1\Big) +\ldots,
\end{eqnarray}

\noindent
where $a_\varphi \equiv M_\varphi/\Lambda$ and $a \equiv M/\Lambda$. Note that in this expression the dependence of the one-loop result on $y_0$ is written explicitly. The presence of this parameter is needed, because due to its renormalization all powers of $\ln\Lambda/\mu$ disappear if $\gamma_c$ is expressed in terms of the bare couplings. In principle, this requirement (i.e. absence of $\ln\Lambda/\mu$ inside $\gamma_c$) allows to find the renormalization of the parameter $y$ in a different way in comparison with Refs. \cite{Juer:1982fb,Juer:1982mp}. The results of these two calculations exactly coincide.

However, in the two-loop approximation we do not write the parameters describing the nonlinear renormalization of the quantum gauge superfield, because we make the calculation for ${\cal F}(V) = V$. Note that besides $y$, similar constants in Eq. (\ref{Nonlinear_Function}) corresponding to the terms of higher orders in $V$
are essential in the two-loop approximation. All these parameters vanish for ${\cal F}(V) = V$. Therefore, if we use this condition, in the final result for the anomalous dimension it is necessary to set $y_0=0$, so that

\begin{eqnarray}\label{Bare_Gamma}
&&\hspace*{-5mm} \gamma_c(\alpha_0,\lambda_0,\xi_0,y_0=0) = \frac{\alpha_0 C_2 (\xi_0-1)}{6\pi} -  \frac{\alpha_0^2}{24\pi^2} C_2^2 \Big(\xi_0^2-1\Big) - \frac{\alpha_0^2 C_2^2}{4\pi^2}\Big( \ln a_\varphi + 1\Big) \nonumber\\
&&\hspace*{-5mm} + \frac{\alpha_0^2 C_2 T(R)}{12\pi^2}\Big( \ln a + 1\Big) +\mbox{higher orders}.
\end{eqnarray}

\section{Anomalous dimension defined in terms of the renormalized couplings}
\hspace*{\parindent}\label{Section_Gamma_Renormalized}

Now, let us calculate the anomalous dimension of the Faddeev--Popov ghosts (standardly) defined in terms of the renormalized couplings. For this purpose we integrate the renormalization group equation

\begin{equation}\label{Gamma_C_RG_Equation_Bare}
\gamma_c(\alpha_0,\lambda_0,\xi_0,y_0) = \left.-\frac{d\ln Z_c}{d\ln\Lambda}\right|_{\alpha,\lambda,\xi,y=\mbox{\scriptsize const}}
\end{equation}

\noindent
and construct the ghost renormalization constant,

\begin{eqnarray}\label{Z_C_Renormalized}
&&\hspace*{-8mm} \ln Z_c = -\frac{\alpha C_2 (\xi-1)}{6\pi} \ln\frac{\Lambda}{\mu} + \frac{\alpha C_2}{\pi} h_1  + \frac{5\alpha y C_2^2(\xi-1)}{4\pi} \ln\frac{\Lambda}{\mu} +  \frac{\alpha^2 C_2^2}{24\pi^2} \Big(\xi^2-1\Big) \ln\frac{\Lambda}{\mu}  \nonumber\\
&&\hspace*{-8mm} - \frac{\alpha^2 C_2^2}{4\pi^2} \Big[\frac{1}{2}\ln^2\frac{\Lambda}{\mu} +\Big(b_{11} -\ln a_\varphi-1\Big)\ln \frac{\Lambda}{\mu} \Big] + \frac{\alpha^2 C_2 T(R)}{12\pi^2}\Big[\frac{1}{2}\ln^2\frac{\Lambda}{\mu}+\Big(b_{12} -\ln a -1\Big)\ln\frac{\Lambda}{\mu} \Big]  \nonumber\\
&&\hspace*{-8mm} - \frac{\alpha^2 C_2^2}{72\pi^2} \Big[\frac{1}{2}(\xi-1)(\xi-2)\ln^2\frac{\Lambda}{\mu} + \Big(4 x_1 - (\xi-1) k_1\Big)\ln\frac{\Lambda}{\mu} \Big] + \frac{\alpha^2 C_2^2}{\pi^2} h_{21} + \frac{\alpha^2 C_2 T(R)}{\pi^2} h_{22}\nonumber\\
&&\hspace*{-8mm} + \ldots\vphantom{\frac{1}{2}}
\end{eqnarray}

\noindent
Note that in the terms coming from the one-loop approximation we so far retain the dependence on the parameter $y$, while in the two-loop contribution we set $y=0$. Again, the finite constants $h_1$, $h_{21}$, and $h_{22}$ appear in the expression (\ref{Z_C_Renormalized}) due to arbitrariness of choosing a subtraction scheme. Using Eqs. (\ref{Alpha_Renormalization}), (\ref{Alpha_Xi_Renormalization}), and (\ref{Y_Renormalization_Copy}) it is possible to rewrite the expression for $\ln Z_c$ in terms of the bare couplings,

\begin{eqnarray}\label{Z_C_Bare}
&&\hspace*{-10mm} \ln Z_c = -\frac{\alpha_0 C_2 (\xi_0-1)}{6\pi} \ln\frac{\Lambda}{\mu} + \frac{\alpha_0 C_2}{\pi} h_1  + \frac{5\alpha_0 y_0 C_2^2 (\xi_0-1)}{4\pi} \ln\frac{\Lambda}{\mu} +  \frac{\alpha_0^2 C_2^2}{24\pi^2} \Big(\xi_0^2-1\Big) \ln\frac{\Lambda}{\mu}  \nonumber\\
&&\hspace*{-10mm} + \frac{\alpha_0^2 C_2^2}{4\pi^2} \Big[\frac{1}{2}\ln^2\frac{\Lambda}{\mu} +\Big(\ln a_\varphi+1 +6h_1\Big)\ln \frac{\Lambda}{\mu} + 6 h_1 b_{11}\Big] - \frac{\alpha_0^2 C_2 T(R)}{12\pi^2}\Big[\frac{1}{2}\ln^2\frac{\Lambda}{\mu}+\Big(\ln a +1 \nonumber\\
&&\hspace*{-10mm} + 6h_1\Big)\ln\frac{\Lambda}{\mu} + 6 h_1 b_{12}\Big] + \frac{\alpha_0^2 C_2^2}{144\pi^2} (\xi_0-1)(\xi_0-2)\ln^2\frac{\Lambda}{\mu} + \frac{\alpha_0^2 C_2^2}{\pi^2} h_{21} + \frac{\alpha_0^2 C_2 T(R)}{\pi^2} h_{22} + \ldots,
\end{eqnarray}

\noindent
where we omit all $y$-dependent terms proportional to $\alpha^2$. Then the anomalous dimension defined in terms of the renormalized couplings is obtained by differentiating Eq. (\ref{Z_C_Bare}) with respect to $\ln\mu$,

\begin{equation}\label{Gamma_C_RG_Equation_Renormalized}
\widetilde\gamma_c(\alpha,\lambda,\xi,y) = \left.\frac{d\ln Z_c}{d\ln\mu}\right|_{\alpha_0,\lambda_0,\xi_0,y_0=\mbox{\scriptsize const}}.
\end{equation}

\noindent
This gives the expression depending on the bare couplings and $\ln\Lambda/\mu$,

\begin{eqnarray}
&& \widetilde\gamma_c(\alpha,\lambda,\xi,y) = \frac{\alpha_0 C_2 (\xi_0-1)}{6\pi} -\frac{5\alpha_0 y_0 C_2^2 (\xi_0-1)}{4\pi} - \frac{\alpha_0^2 C_2^2}{4\pi^2}\Big(\ln\frac{\Lambda}{\mu} + \ln a_\varphi + 1 + 6 h_1\Big) \nonumber\\
&& + \frac{\alpha_0^2 C_2 T(R)}{12\pi^2}\Big(\ln\frac{\Lambda}{\mu} + \ln a + 1 + 6h_1\Big)
-\frac{\alpha_0^2 C_2^2}{24\pi^2}\Big(\xi_0^2-1\Big) - \frac{\alpha_0^2 C_2^2}{72\pi^2} (\xi_0-1)(\xi_0-2) \ln\frac{\Lambda}{\mu}\qquad\nonumber\\
&& + \ldots\vphantom{\frac{1}{2}}
\end{eqnarray}

\noindent
Certainly, it is necessary to express the result in terms of the renormalized couplings again using Eqs. (\ref{Alpha_Renormalization}), (\ref{Alpha_Xi_Renormalization}), and (\ref{Y_Renormalization_Copy}),

\begin{eqnarray}
&&\hspace*{-8mm} \widetilde\gamma_c(\alpha,\lambda,\xi,y) =  \frac{\alpha C_2 (\xi-1)}{6\pi} -\frac{5\alpha y C_2^2 (\xi-1)}{4\pi} - \frac{\alpha^2 C_2^2}{4\pi^2}\Big(\ln a_\varphi + 1 + 6h_1 - b_{11}\Big) \nonumber\\
&&\hspace*{-8mm} + \frac{\alpha^2 C_2 T(R)}{12\pi^2}\Big( \ln a + 1 + 6h_1-b_{12}\Big) -\frac{\alpha^2 C_2^2}{24\pi^2}\Big(\xi^2-1\Big) + \frac{\alpha^2 C_2^2}{72\pi^2}\Big(4 x_1 - (\xi-1) k_1\Big)+\ldots
\end{eqnarray}

\noindent
Setting in this expression $y=0$ we obtain the final result for the two-loop Faddeev--Popov ghost anomalous dimension defined in terms of the renormalized coupling constant for the case when all renormalized parameters describing the nonlinear renormalization vanish,

\begin{eqnarray}\label{Renormalized_Gamma}
&& \widetilde\gamma_c(\alpha,\lambda,\xi,y=0) =  \frac{\alpha C_2 (\xi-1)}{6\pi} - \frac{\alpha^2 C_2^2}{4\pi^2}\Big(\ln a_\varphi + 1 + 6h_1 - b_{11}\Big) + \frac{\alpha^2 C_2 T(R)}{12\pi^2}\Big( \ln a + 1 \qquad\nonumber\\
&& + 6h_1-b_{12}\Big) -\frac{\alpha^2 C_2^2}{24\pi^2}\Big(\xi^2-1\Big) + \frac{\alpha^2 C_2^2}{72\pi^2}\Big(4 x_1 - (\xi-1) k_1\Big) + \mbox{higher orders}.\vphantom{\frac{1}{2}}
\end{eqnarray}

\noindent
Unlike the anomalous dimension defined in terms of the bare couplings, this expression depends on the finite constants and is, therefore, scheme-dependent.

Note that after the formal substitution $\alpha_0\to \alpha$, $\xi_0\to \xi$ the function (\ref{Bare_Gamma}) coincides with the function (\ref{Renormalized_Gamma}) in the $\mbox{HD}+\mbox{MSL}$ scheme (see \cite{Kataev:2013eta,Kazantsev:2017fdc,Kataev:2017qvk,Stepanyantz:2017sqg}).\footnote{Certainly, this can be easily proved in all orders by repeating the argumentation of Ref. \cite{Kataev:2013eta}.} This renormalization prescription means that the regularization is made by the help of the higher covariant derivative method, and only powers of $\ln\Lambda/\mu$ are included into the renormalization constants, so that all finite constants should be set to 0, in particular, $b_{11}=b_{12}=0$, $h_1=0$, $k_1=0$, $x_1=0$.

\section*{Conclusion}
\hspace*{\parindent}

In this paper the two-loop anomalous dimension of the Faddeev--Popov ghosts is obtained for the general renormalizable ${\cal N}=1$ supersymmetric gauge theory, regularized by higher covariant derivatives. We demonstrate that for doing this calculation it is very important that the quantum gauge superfield is renormalized in a nonlinear way. Without this nonlinear renormalization the renormalization group equations (\ref{Gamma_C_RG_Equation_Bare}) or (\ref{Gamma_C_RG_Equation_Renormalized}) for the ghost renormalization constant are not satisfied. However, the nonlinear term obtained in Refs. \cite{Juer:1982fb,Juer:1982mp} amends the situation. Note that the coefficient of the nonlinear term needed for deriving the anomalous dimension of the Faddeev--Popov ghosts exactly coincides with the one found in \cite{Juer:1982fb,Juer:1982mp} from a different calculation.

The anomalous dimension of the Faddeev--Popov ghosts obtained in this paper can be used for making the two-loop verification of the non-renormalization theorem for the triple ghost-gauge vertices \cite{Stepanyantz:2016gtk} (for this purpose it is also necessary to know the two-loop renormalization of the coupling constant and the two-loop linear part of the quantum gauge superfield renormalization with the considered regularization). Another important application is checking the new form of the exact NSVZ $\beta$-function (which relates the $\beta$-function to the anomalous dimensions of the Faddeev--Popov ghosts, of the quantum gauge superfield and of the matter superfields) for the three-loop diagrams which include ghost loops. This allows to verify the term containing $\gamma_c$ in Eq. (\ref{NSVZ_Equation_New}) in such an approximation where the scheme dependence is essential.

\section*{Acknowledgements}
\hspace*{\parindent}

A.K. and K.S. are very grateful to A.L.Kataev for valuable discussions.

The work of A.K., N.M., I.S., M.S. was supported by the Foundation for the advancement of theoretical physics and mathematics ``BASIS'', grants No. 17-11-120 (A.K., M.S.),
17-22-222 (N.M.), 17-22-2205 (I.S.).

\appendix

\section*{Appendix}

\section{Nontrivial contributions to the function $G_c$}
\hspace*{\parindent}\label{Appendix_Superdiagrams_For_Gc}

In this appendix we present expressions for various contributions to the function $G_c$ defined by Eq. (\ref{G_C_Definition}) from the superdiagrams depicted in Fig. \ref{Figure_Diagrams} in the limit of the vanishing external momentum. Some of them vanish, namely,

\begin{eqnarray}
&& \Delta G_c^{(1)} = 0;\qquad\ \Delta G_c^{(3)} = 0;\qquad\ \Delta G_c^{(4)} = 0;\qquad\ \Delta G_c^{(5)} = 0;\qquad\ \Delta G_c^{(6)} = 0;\qquad\nonumber\\
&& \Delta G_c^{(7)} = 0;\qquad\ \Delta G_c^{(9)} = 0;\qquad\ \Delta G_c^{(10)} = 0;\qquad \Delta G_c^{(11)} = 0;\qquad \Delta G_c^{(14)} = 0.\qquad
\end{eqnarray}

\noindent
Contributions of the other diagrams to the function $G_c$ are presented in the Euclidean space after the Wick rotation. Certainly, in the limit of the vanishing external momentum they are written formally, because the corresponding integrals diverge. However, by the help of Eq. (\ref{Gamma_C_General}) from these formal expressions one can construct the expression for the anomalous dimension $\gamma_c$, which is well defined.

\begin{eqnarray}\label{Diagram1}
&& \Delta G_c^{(2)} = \frac{e_0^2 C_2}{3} (\xi_0-1) \int \frac{d^4k}{(2\pi)^4} \frac{1}{k^4 R_k} -\frac{5 e_0^2 y_0}{2} C_2^2 (\xi_0-1) \int \frac{d^4k}{(2\pi)^4} \frac{1}{k^4 R_k};\\
&& \Delta G_c^{(8)} = \frac{e_0^4 C_2^2}{4} \xi_0(\xi_0-1) \int \frac{d^4k}{(2\pi)^4} \frac{d^4l}{(2\pi)^4} \frac{1}{k^4 R_k l^4 R_l};\qquad\\
\label{Diagram12}
&& \Delta G_c^{(12)} = -\frac{e_0^4 C_2^2}{6} (\xi_0-1) \int \frac{d^4k}{(2\pi)^4} \frac{d^4l}{(2\pi)^4} \frac{1}{R_k R_l} \left( \frac{\xi_0+1}{k^2 l^4 (k+l)^2} + \frac{\xi_0-1}{2 k^4 l^4}\right);\qquad\\
\label{Diagram13}
&& \Delta G_c^{(13)} = -\frac{e_0^4 C_2^2}{18} (\xi_0-1)^2 \int \frac{d^4k}{(2\pi)^4} \frac{d^4l}{(2\pi)^4} \frac{1}{k^4 R_k l^4 R_l};\\
\label{Diagram15}
&& \Delta G_c^{(15)} = \frac{2e_0^4 C_2}{3} \int \frac{d^4k}{(2\pi)^4} \frac{1}{k^4 R_k^2}\Big(C_2 f(k/\Lambda) + C_2 g(\xi_0,k/\Lambda) + T(R) h(k/\Lambda)\Big).\qquad
\end{eqnarray}

\noindent
The functions $f(k/\Lambda)$, $g(\xi_0,k/\Lambda)$, and $h(k/\Lambda)$ entering Eq. (\ref{Diagram15}) have been calculated in Ref. \cite{Kazantsev:2017fdc}. For completeness, we also present them in Appendix \ref{Appendix_Loop_Integrals}.

\section{One-loop polarization operator insertion}
\hspace*{\parindent}\label{Appendix_Loop_Integrals}

Let us present explicit expressions for the functions $f(k/\Lambda)$, $g(\xi_0,k/\Lambda)$, and $h(k/\Lambda)$. Following Ref. \cite{Kazantsev:2017fdc},\footnote{Note that in this paper we use the gauge fixing term with the regulator function $K=R$. The expression (\ref{G_Function}) has been obtained from the corresponding result of Ref. \cite{Kazantsev:2017fdc} after this substitution.} the gauge dependent part of the one-loop polarization operator is included into the function

\begin{eqnarray}\label{G_Function}
&& g(\xi_0,k/\Lambda) = \int \frac{d^4l}{(2\pi)^4} \Bigg[\frac{(\xi_0-1)}{2l^4 R_l} \Big(R_{k+l} - \frac{2}{3} R_k\Big) - \frac{(\xi_0-1)}{2 l^2 R_l (k+l)^4 R_{k+l}} (k_\mu R_k + l_\mu R_l)^2\qquad\nonumber\\
&& - \frac{(\xi_0-1)^2 k^2 R_k^2}{4 l^4 R_l (k+l)^4 R_{k+l}} l^\mu (k+l)_\mu \Bigg].
\end{eqnarray}

\noindent
It is convenient to present the remaining functions $f$ and $h$ in the form

\begin{equation}
f(k/\Lambda) = f_1(k/\Lambda) + f_2(k/\Lambda);\qquad h(k/\Lambda) = h_1(k/\Lambda) + h_2(k/\Lambda),
\end{equation}

\noindent
where

\begin{eqnarray}
&&\hspace*{-5mm} f_1(k/\Lambda) \equiv - \frac{3}{2} \int \frac{d^4l}{(2\pi)^4} \left(\frac{1}{l^2(l+k)^2} -\frac{1}{(l^2+M_{\varphi}^2)((l+k)^2+M_{\varphi}^2)}\right);\\
&&\hspace*{-5mm} f_2(k/\Lambda) = \int\frac{d^4l}{(2\pi)^4} \Biggl[ \frac{R_{l}-R_{k}}{R_{l}l^2}\left(\frac{1}{(l+k)^2} -\frac{1}{l^2-k^2}\right) +\frac{2}{R_l \big((l+k)^2-l^2\big)}\left(\frac{R_{l+k}-R_{l}}{(l+k)^2-l^2}\right.\nonumber\\
&&\hspace*{-5mm} \left. -\frac{R'_{l}}{\Lambda^2}\right) - \frac{1}{R_{l}R_{l+k}} \left(\frac{R_{l+k}-R_{l}}{(l+k)^2-l^2}\right)^2   - \frac{2 R_{k}k^2}{l^2(l+k)^2R_{l}R_{l+k}} \left(\frac{R_{l+k}-R_{k}}{(l+k)^2-k^2}\right) - \frac{l_\mu k^\mu R_{k}}{l^2R_{l}(l+k)^2R_{l+k}}\nonumber\\
&&\hspace*{-5mm} \times\left(\frac{R_{l+k}-R_{l}}{(l+k)^2-l^2}\right) + \frac{2 l_\mu k^\mu}{l^2R_{l}R_{l+k}} \left(\frac{R_{l+k}-R_{k}}{(l+k)^2-k^2}\right) \left(\frac{R_{l+k}-R_{l}}{(l+k)^2-l^2}\right) - \frac{2 k^2}{(l+k)^2R_{l}R_{l+k}}\nonumber\\
&&\hspace*{-5mm} \times \left(\frac{R_{l}-R_{k}}{l^2-k^2}\right)^2 -\frac{k^2 l_\mu (l+k)^\mu}{l^2(l+k)^2R_{l}R_{l+k}} \left(\frac{R_{l}-R_{k}}{l^2-k^2}\right) \left(\frac{R_{l+k}-R_{k}}{(l+k)^2-k^2}\right) +\frac{2k^2}{\big((l+k)^2-k^2\big) l^2 R_l}\nonumber\\
&&\hspace*{-5mm} \times \left(\frac{R_{l+k}-R_{k}}{(l+k)^2-k^2} -\frac{R'_{k}}{\Lambda^2}\right) -\frac{2 l_\mu k^\mu}{l^2R_{l}} \left(\frac{R_{l}}{\left(l^2-(l+k)^2\right) \left(l^2-k^2\right)}+\frac{R_{l+k}}{\left((l+k)^2-l^2\right) \left((l+k)^2-k^2\right)}\right.\nonumber\\
&&\hspace*{-5mm} \left.+\frac{R_{k}}{\left(k^2-l^2\right)\left(k^2-(l+k)^2\right)}\right) - \frac{1}{2\big((l+k)^2-l^2\big)}\left(\frac{2 R_{l+k} R'_{l+k} (l+k)^2}{\Lambda^2\big((l+k)^2 R_{l+k}^2+M_\varphi^2\big)} -\frac{2 R_l R'_l l^2}{\Lambda^2\big(l^2 R_l^2 + M_\varphi^2\big)}\right.\nonumber\\
&&\hspace*{-5mm} \left. - \frac{1}{(l+k)^2 + M_\varphi^2} + \frac{1}{l^2+M_\varphi^2} +\frac{R_{l+k}^2}{(l+k)^2 R_{l+k}^2 + M_\varphi^2}
- \frac{R_l^2}{l^2 R_l^2+M_\varphi^2} \right)\Biggr];\\
&&\hspace*{-5mm} h_1(k/\Lambda) \equiv \frac{1}{2} \int \frac{d^4l}{(2\pi)^4} \left(\frac{1}{l^2(l+k)^2} -\frac{1}{(l^2+M^2)((l+k)^2+M^2)}\right);\\
&&\hspace*{-5mm} h_2(k/\Lambda) = \int \frac{d^4l}{(2\pi)^4} \frac{1}{\big((k+l)^2-l^2\big)}\left(\vphantom{\frac{1}{2}}\right. - \frac{M^2 F'_{k+l}}{\Lambda^2 F_{k+l} \big((k+l)^2 F_{k+l}^2 + M^2\big)} + \frac{M^2 F'_{l}}{\Lambda^2 F_l \big(l^2 F_{l}^2 + M^2\big)} \nonumber\\
&&\hspace*{-5mm} -\frac{1}{2\big((k+l)^2+M^2\big)} + \frac{1}{2\big(l^2+M^2\big)} + \frac{F_{k+l}^2}{2\big((k+l)^2 F_{k+l}^2 + M^2\big)} - \frac{F_l^2}{2\big(l^2 F_l^2+M^2\big)} \left.\vphantom{\frac{1}{2}}\right)
\end{eqnarray}

\noindent
with

\begin{equation}
R_q' \equiv \frac{\partial R(q^2/\Lambda^2)}{\partial(q^2/\Lambda^2)};\qquad F_q' \equiv \frac{\partial F(q^2/\Lambda^2)}{\partial(q^2/\Lambda^2)}.
\end{equation}

\noindent
Let us also recall that $M_\varphi$ denotes the mass of the Pauli--Villars superfields which compensate the one-loop divergences of diagrams with a loop of the quantum gauge superfield and of the ghosts. Similarly, $M$ is the mass of the Pauli--Villars superfields which cancel the one-loop divergences from a matter loop.

According to Refs. \cite{Soloshenko:2003nc,Soloshenko:2002np} (see also Ref. \cite{Kataev:2017qvk}), for the regulator $R(x) = 1+ x^n$

\begin{eqnarray}
&&\hspace*{-10mm} \int \frac{d^4k}{(2\pi)^4} \frac{1}{k^4} \frac{d}{d\ln\Lambda} \Big[\frac{f_1(k/\Lambda)}{R_k^2} + \frac{3}{16\pi^2 R_k}\Big(\ln\frac{\Lambda}{\mu} + b_{11}\Big) \Big] =
\frac{3}{128\pi^4}\Big(\ln\frac{\Lambda}{\mu} + b_{11} - \ln a_\varphi -1 \Big);\\
&&\hspace*{-10mm} \int \frac{d^4k}{(2\pi)^4} \frac{1}{k^4} \frac{d}{d\ln\Lambda} \Big[\frac{h_1(k/\Lambda)}{R_k^2} - \frac{1}{16\pi^2 R_k}\Big(\ln\frac{\Lambda}{\mu} + b_{12}\Big) \Big] =
-\frac{1}{128\pi^4}\Big(\ln\frac{\Lambda}{\mu} + b_{12} - \ln a -1 \Big),
\end{eqnarray}

\noindent
where $a_\varphi = M_\varphi/\Lambda$ and $a=M/\Lambda$.

The remaining integrals containing the functions $f_2(k/\Lambda)$ and $h_2(k/\Lambda)$ can be calculated repeating the argumentation of Refs.
\cite{Kazantsev:2017fdc,Kataev:2017qvk}. Namely, the derivatives with respect to $\ln\Lambda$ are converted into the derivatives with respect to $\ln k$. Then the integrals can be taken, the results being proportional to $f_2(0)$ and $h_2(0)$. These values appear to be zero, see Ref. \cite{Kazantsev:2017fdc} for details. Thus, we obtain

\begin{eqnarray}
&& \int \frac{d^4k}{(2\pi)^4} \frac{d}{d\ln\Lambda} \frac{f_2(k/\Lambda)}{k^4 R_k^2} = \frac{1}{8\pi^2} f_2(0) = 0;\\
&& \int \frac{d^4k}{(2\pi)^4} \frac{d}{d\ln\Lambda} \frac{h_2(k/\Lambda)}{k^4 R_k^2} = \frac{1}{8\pi^2} h_2(0) = 0.
\end{eqnarray}

\section{Calculation of the anomalous dimension}
\hspace*{\parindent}\label{Appendix_Calculation}

Let us describe in detail the calculation of integrals giving the anomalous dimension $\gamma_c$. Substituting the expressions (\ref{Alpha_Renormalization}), (\ref{Alpha_Xi_Renormalization}), and (\ref{Y_Renormalization_Copy}) into Eq. (\ref{Gamma_C_Integral}), $\gamma_c$ can be written as

\begin{eqnarray}
&&\hspace*{-7mm} \gamma_c(\alpha_0,\lambda_0,\xi_0,y_0) = \frac{d}{d\ln\Lambda}\Bigg\{ 4\pi C_2  \int \frac{d^4k}{(2\pi)^4} \frac{\alpha}{k^4 R_k}\Bigg[(\xi-1) \Big(\frac{1}{3} - \frac{5}{2} y C_2\Big) + \frac{8\pi\alpha}{3 R_k}\Bigg(C_2 \Big[f(k/\Lambda)\nonumber\\
&&\hspace*{-7mm} + \frac{3R_k}{16\pi^2} \Big(\ln\frac{\Lambda}{\mu} + b_{11}\Big) \Big] + T(R)\, \Big[h(k/\Lambda) - \frac{R_k}{16\pi^2}\Big(\ln\frac{\Lambda}{\mu}+ b_{12}\Big)\Big]\Bigg) \Bigg]  + 4\pi^2 C_2^2 \alpha^2 (\xi-1) \nonumber\\
&&\hspace*{-7mm} \times  \int \frac{d^4k}{(2\pi)^4} \frac{d^4l}{(2\pi)^4} \frac{1}{R_k R_l}\Bigg[\frac{(5\xi+8)}{9k^4 l^4} - \frac{4(\xi+1)}{3 k^4 l^2 (k+l)^2} \Bigg]
+ \frac{1}{9}\alpha^2 C_2^2 \int \frac{d^4k}{(2\pi)^4} \frac{1}{k^4 R_k} \Bigg[\, 4 x_1 - (\xi-1) k_1
\nonumber\\
&& \hspace*{-7mm}  + (\xi-1)(\xi-2) \ln\frac{\Lambda}{\mu} \Bigg]\Bigg\}_{\alpha,\xi,y=\mbox{\scriptsize const}} +\ldots,
\end{eqnarray}

\noindent
where dots denote the two-loop terms containing parameters of the nonlinear renormalization and the terms of higher orders. The integral containing the function $f(k/\Lambda)$ has been found in Ref. \cite{Kataev:2017qvk}. The integral containing the function $h(k/\Lambda)$ can be calculated similarly. The main ideas of the calculation are described in Appendix \ref{Appendix_Loop_Integrals}. Also we will use the identity

\begin{equation}\label{Integral_Identity}
I\equiv \frac{d}{d\ln\Lambda} \int \frac{d^4k}{(2\pi)^4} \frac{d^4l}{(2\pi)^4} \frac{1}{R_k R_l}\Big(\frac{1}{k^4 l^2 (k+l)^2} - \frac{1}{2 k^4 l^4}\Big) = \frac{1}{128\pi^4}.
\end{equation}

\noindent
To prove this equation we note that the left hand side can be presented in the form

\begin{eqnarray}
&& I = - \int \frac{d^4k}{(2\pi)^4} \frac{d^4l}{(2\pi)^4}\, \frac{\partial}{\partial l^\mu} \left(\frac{l^\mu}{2 l^4 k^2 (k+l)^2}\frac{d}{d\ln\Lambda} \frac{1}{R_{k+l} R_l}\right)\nonumber\\
&&\qquad\qquad\quad - \frac{d}{d\ln\Lambda}\int \frac{d^4k}{(2\pi)^4} \frac{d^4l}{(2\pi)^4}\, \frac{l^\mu}{\Lambda^2 l^4 k^2 (k+l)^2} \left(\frac{(k+l)_\mu R_{k+l}'}{R_{k+l}^2 R_l} + \frac{l_\mu R_l'}{R_l^2 R_{k+l}} \right).\qquad
\end{eqnarray}

\noindent
The first integral can be reduced to a surface integral and easily calculated. The derivative of the second integral with respect to $\ln\Lambda$ vanishes. Really, this integral is convergent in both the ultraviolet and infrared regions and is, therefore, a finite constant independent of $\Lambda$. Thus, the considered integral can be rewritten as

\begin{equation}
I = - \frac{1}{16\pi^4} \oint\limits_{S^3_\infty + S^3_\varepsilon} dS_\mu  \int \frac{d^4k}{(2\pi)^4} \left(\frac{l^\mu}{2 l^4 k^2 (k+l)^2}\frac{d}{d\ln\Lambda} \frac{1}{R_{k+l} R_l}\right),
\end{equation}

\noindent
where the surface integration is done over the sphere $S^3_\infty$ of the infinitely large radius in the space of the Euclidean momentum $l^\mu$ and over the infinitely small sphere $S^3_\varepsilon$ surrounding the point $l_\mu=0$. (In our notation the normal vector to the sphere $S^3_\varepsilon$ is inward-pointing.) The integral over $S^3_\infty$ evidently vanishes due to the higher derivative regulators $R_l$ and $R_{k+l}$. Calculating the remaining integral over $S^3_\varepsilon$ we obtain

\begin{equation}
I = \frac{1}{16\pi^2} \frac{d}{d\ln\Lambda} \int\frac{d^4k}{(2\pi)^4} \frac{1}{k^4 R_k} = \frac{1}{128\pi^4}.
\end{equation}

By the help of Eq. (\ref{Integral_Identity}) the expression for the anomalous dimension $\gamma_c$ can be presented in the form

\begin{eqnarray}\label{Gamma_C_2}
&&\hspace*{-5mm} \gamma_c(\alpha_0,\lambda_0,\xi_0,y_0) = \frac{\alpha C_2 (\xi-1)}{6\pi} - \frac{5\alpha y C_2^2(\xi-1)}{4\pi}  -  \frac{\alpha^2}{24\pi^2} C_2^2 \Big(\xi^2-1\Big) + \frac{\alpha^2 C_2^2}{4\pi^2}\Big(\ln\frac{\Lambda}{\mu}+ b_{11} - \ln a_\varphi \nonumber\\
&&\hspace*{-5mm} -1\Big) - \frac{\alpha^2 C_2 T(R)}{12\pi^2}\Big(\ln\frac{\Lambda}{\mu}+ b_{12} - \ln a -1\Big) + \frac{\alpha^2}{72\pi^2} C_2^2 \Big(4 x_1 - (\xi-1) k_1 \Big)
- \frac{4\pi^2}{9} \alpha^2 C_2^2 (\xi-1) \nonumber\\
&&\hspace*{-5mm} \times (\xi-2) \frac{d}{d\ln\Lambda} \left(\int\frac{d^4l}{(2\pi)^4} \frac{d^4k}{(2\pi)^4} \frac{1}{k^4 R_k l^4 R_l} - \frac{1}{4\pi^2} \ln\frac{\Lambda}{\mu} \int \frac{d^4k}{(2\pi)^4} \frac{1}{k^4 R_k}\right) +\ldots.
\end{eqnarray}

\noindent
To calculate the remaining integral we note that

\begin{eqnarray}
&& \frac{d}{d\ln\Lambda} \left(\int\frac{d^4l}{(2\pi)^4} \frac{d^4k}{(2\pi)^4} \frac{1}{k^4 R_k l^4 R_l} - \frac{1}{4\pi^2} \ln\frac{\Lambda}{\mu} \int \frac{d^4k}{(2\pi)^4} \frac{1}{k^4 R_k}\right) \nonumber\\
&& = \frac{d}{d\ln\Lambda} \Bigg[\Big(\int \frac{d^4k}{(2\pi)^4}\frac{1}{k^4 R_k} - \frac{1}{8\pi^2}\ln\frac{\Lambda}{\mu}\Big)^2 - \frac{1}{64\pi^4} \ln^2 \frac{\Lambda}{\mu}\Bigg] = - \frac{1}{32\pi^4} \ln \frac{\Lambda}{\mu},\qquad
\end{eqnarray}

\noindent
because

\begin{equation}
\frac{d}{d\ln\Lambda} \Big(\int \frac{d^4k}{(2\pi)^4} \frac{1}{k^4 R_k} - \frac{1}{8\pi^2}\ln\frac{\Lambda}{\mu}\Big)^2 \equiv \lim\limits_{p\to 0} \frac{d}{d\ln\Lambda} \Big(\int \frac{d^4k}{(2\pi)^4} \frac{1}{k^2 (k+p)^2 R_k} - \frac{1}{8\pi^2}\ln\frac{\Lambda}{\mu}\Big)^2 = 0.
\end{equation}

\noindent
Therefore, the expression for the anomalous dimension can be written in the form

\begin{eqnarray}
&&\hspace*{-7mm} \gamma_c(\alpha_0,\lambda_0,\xi_0,y_0) = \frac{\alpha C_2 (\xi-1)}{6\pi} - \frac{5\alpha y C_2^2(\xi-1)}{4\pi} -  \frac{\alpha^2}{24\pi^2} C_2^2 (\xi^2-1)
+ \frac{\alpha^2 C_2^2}{4\pi^2}\Big(\ln\frac{\Lambda}{\mu}  \nonumber\\
&&\hspace*{-7mm} + b_{11} - \ln a_\varphi -1\Big) - \frac{\alpha^2 C_2 T(R)}{12\pi^2}\Big(\ln\frac{\Lambda}{\mu}+ b_{12} - \ln a -1\Big)  + \frac{\alpha^2 C_2^2}{72\pi^2} \Big( (\xi-1)(\xi-2)\ln\frac{\Lambda}{\mu}   \nonumber\\
&&\hspace*{-7mm} + 4 x_1 -  (\xi-1) k_1 \Big) +\ldots\vphantom{\frac{1}{2}}
\end{eqnarray}

\noindent
Let us recall that this anomalous dimension is defined in terms of the bare couplings, so that the right hand side should be expressed in terms of the bare couplings by the help of Eqs. (\ref{Alpha_Renormalization}), (\ref{Alpha_Xi_Renormalization}), and (\ref{Y_Renormalization_Copy}). This gives the result for the considered renormalization group function,

\begin{eqnarray}
&& \gamma_c(\alpha_0,\lambda_0,\xi_0,y_0) = \frac{\alpha_0 C_2 (\xi_0-1)}{6\pi} - \frac{5\alpha_0 y_0 C_2^2(\xi_0-1)}{4\pi}  \nonumber\\
&&\qquad\qquad -  \frac{\alpha_0^2}{24\pi^2} C_2^2 \Big(\xi_0^2-1\Big) - \frac{\alpha_0^2 C_2^2}{4\pi^2}\Big( \ln a_\varphi + 1\Big) + \frac{\alpha_0^2 C_2 T(R)}{12\pi^2}\Big( \ln a + 1\Big) +\ldots \qquad
\end{eqnarray}


\begin{thebibliography}{100}

%\cite{Grisaru:1979wc}
\bibitem{Grisaru:1979wc}
  M.~T.~Grisaru, W.~Siegel and M.~Rocek,
  %``Improved Methods for Supergraphs,''
  Nucl.\ Phys.\ B {\bf 159} (1979) 429.
  %%CITATION = NUPHA,B159,429;%%

%\cite{Novikov:1983uc}
\bibitem{Novikov:1983uc}
  V.~A.~Novikov, M.~A.~Shifman, A.~I.~Vainshtein and V.~I.~Zakharov,
  %``Exact Gell-Mann-Low Function of Supersymmetric Yang-Mills Theories from Instanton Calculus,''
  Nucl.\ Phys.\ B {\bf 229} (1983) 381.
  %%CITATION = NUPHA,B229,381;%%

%\cite{Jones:1983ip}
\bibitem{Jones:1983ip}
  D.~R.~T.~Jones,
  %``More on the Axial Anomaly in Supersymmetric {Yang-Mills} Theory,''
  Phys.\ Lett.\ B {\bf 123} (1983) 45.
  %%CITATION = PHLTA,B123,45;%%

%\cite{Novikov:1985rd}
\bibitem{Novikov:1985rd}
  V.~A.~Novikov, M.~A.~Shifman, A.~I.~Vainshtein and V.~I.~Zakharov,
  %``Beta Function in Supersymmetric Gauge Theories: Instantons Versus Traditional Approach,''
  Phys.\ Lett.\ B {\bf 166} (1986) 329; Sov.\ J.\ Nucl.\ Phys.\  {\bf 43}
(1986) 294; [Yad.\ Fiz.\  {\bf 43} (1986) 459.]
  %%CITATION = PHLTA,B166,329;%%

%\cite{Shifman:1986zi}
\bibitem{Shifman:1986zi}
  M.~A.~Shifman and A.~I.~Vainshtein,
  %``Solution of the Anomaly Puzzle in SUSY Gauge Theories and the Wilson Operator Expansion,''
  Nucl.\ Phys.\ B {\bf 277}  (1986) 456; Sov.\ Phys.\ JETP {\bf 64} (1986) 428;
[Zh.\ Eksp.\ Teor.\ Fiz.\  {\bf 91}  (1986) 723.]
  %%CITATION = NUPHA,B277,456;%%

%\cite{Shifman:1999mv}
\bibitem{Shifman:1999mv}
  M.~A.~Shifman and A.~I.~Vainshtein,
  %``Instantons versus supersymmetry: Fifteen years later,''
  In *Shifman, M.A.: ITEP lectures on particle physics and field theory, vol. 2* 485-647
  [hep-th/9902018].
  %%CITATION = HEP-TH/9902018;%%

%\cite{Shifman:1999kf}
\bibitem{Shifman:1999kf}
  M.~A.~Shifman,
  %``Exact results in gauge theories: Putting supersymmetry to work. The 1999 Sakurai Prize Lecture,''
  Int.\ J.\ Mod.\ Phys.\ A {\bf 14} (1999) 5017.
  %doi:10.1142/S0217751X99002372
  %[hep-th/9906049].
  %%CITATION = doi:10.1142/S0217751X99002372;%%

%\cite{Shifman:2018yxh}
\bibitem{Shifman:2018yxh}
  M.~Shifman,
  %``Supersymmetric Tools in Yang-Mills Theories at Strong Coupling: the Beginning of a Long Journey,''
  Int.\ J.\ Mod.\ Phys.\ A {\bf 33} (2018) no.12,  1830009.
  %doi:10.1142/S0217751X18300090
  %[arXiv:1804.01191 [hep-th]].
  %%CITATION = doi:10.1142/S0217751X18300090;%%

%\cite{Kutasov:2004xu}
\bibitem{Kutasov:2004xu}
  D.~Kutasov and A.~Schwimmer,
  %``Lagrange multipliers and couplings in supersymmetric field theory,''
  Nucl.\ Phys.\ B {\bf 702} (2004) 369.
  %[hep-th/0409029].
  %%CITATION = HEP-TH/0409029;%%

%\cite{Kataev:2014gxa}
\bibitem{Kataev:2014gxa}
  A.~L.~Kataev and K.~V.~Stepanyantz,
  %``The NSVZ $\beta$-function in supersymmetric theories with different regularizations and renormalization prescriptions,''
  Theor.\ Math.\ Phys.\  {\bf 181} (2014) 1531.
  %[arXiv:1405.7598 [hep-th]].
  %%CITATION = ARXIV:1405.7598;%%

%\cite{Siegel:1979wq}
\bibitem{Siegel:1979wq}
  W.~Siegel,
  %``Supersymmetric Dimensional Regularization via Dimensional Reduction,''
  Phys.\ Lett.\ B {\bf 84} (1979) 193.
  %%CITATION = PHLTA,B84,193;%%

%\cite{Siegel:1980qs}
\bibitem{Siegel:1980qs}
  W.~Siegel,
  %``Inconsistency of Supersymmetric Dimensional Regularization,''
  Phys.\ Lett.\ B {\bf 94} (1980) 37.
  %%CITATION = PHLTA,B94,37;%%

%\cite{Jack:1996vg}
\bibitem{Jack:1996vg}
  I.~Jack, D.~R.~T.~Jones and C.~G.~North,
  %``N=1 supersymmetry and the three loop gauge Beta function,''
  Phys.\ Lett.\ B {\bf 386} (1996) 138.

%\cite{Jack:1996cn}
\bibitem{Jack:1996cn}
  I.~Jack, D.~R.~T.~Jones and C.~G.~North,
  %``Scheme dependence and the NSVZ Beta function,''
  Nucl.\ Phys.\ B {\bf 486} (1997) 479.
  %%CITATION = HEP-PH/9609325;%%

%\cite{Jack:1998uj}
\bibitem{Jack:1998uj}
  I.~Jack, D.~R.~T.~Jones and A.~Pickering,
  %``The Connection between DRED and NSVZ,''
  Phys.\ Lett.\ B {\bf 435} (1998) 61.
  %%CITATION = HEP-PH/9805482;%%

%\cite{Slavnov:1971aw}
\bibitem{Slavnov:1971aw}
  A.~A.~Slavnov,
  %``Invariant regularization of nonlinear chiral theories,''
  Nucl.\ Phys.\ B {\bf 31} (1971) 301.
  %%CITATION = NUPHA,B31,301;%%

%\cite{Slavnov:1972sq}
\bibitem{Slavnov:1972sq}
  A.~A.~Slavnov,
  %``Invariant regularization of gauge theories,''
  Theor.Math.Phys. {\bf 13} (1972) 1064
   [Teor.\ Mat.\ Fiz.\  {\bf 13} (1972) 174].
  %%CITATION = TMFZA,13,174;%%

%\cite{Slavnov:1977zf}
\bibitem{Slavnov:1977zf}
  A.~A.~Slavnov,
  %``The Pauli-Villars Regularization for Nonabelian Gauge Theories,''
  Theor.\ Math.\ Phys.\ {\bf 33} (1977) 977
   [Teor.\ Mat.\ Fiz.\  {\bf 33} (1977) 210].
  %%CITATION = TMFZA,33,210;%%

%\cite{Krivoshchekov:1978xg}
\bibitem{Krivoshchekov:1978xg}
  V.~K.~Krivoshchekov,
  %``Invariant Regularizations for Supersymmetric Gauge Theories,''
  Theor.\ Math.\ Phys.\ {\bf 36} (1978) 745
 [Teor.\ Mat.\ Fiz.\  {\bf 36} (1978) 291].
 %%CITATION = TMFZA,36,291;%%

%\cite{West:1985jx}
\bibitem{West:1985jx}
  P.~C.~West,
  %``Higher Derivative Regulation Of Supersymmetric Theories,''
  Nucl.\ Phys.\ B {\bf 268} (1986) 113.
  %%CITATION = NUPHA,B268,113;%%

%\cite{Kataev:2013eta}
\bibitem{Kataev:2013eta}
  A.~L.~Kataev and K.~V.~Stepanyantz,
  %``NSVZ scheme with the higher derivative regularization for $\mathcal{N} =$ 1 SQED,''
  Nucl.\ Phys.\ B {\bf 875} (2013) 459.
  %%CITATION = ARXIV:1305.7094;%%

%\cite{Kataev:2013csa}
\bibitem{Kataev:2013csa}
  A.~L.~Kataev and K.~V.~Stepanyantz,
  %``Scheme independent consequence of the NSVZ relation for $N=1$ SQED with $N_f$ flavors,''
  Phys.\ Lett.\ B {\bf 730} (2014) 184
  %[arXiv:1311.0589 [hep-th]].
  %%CITATION = ARXIV:1311.0589;%%


%\cite{Stepanyantz:2016gtk}
\bibitem{Stepanyantz:2016gtk}
  K.~V.~Stepanyantz,
  %``Non-renormalization of the $V\bar cc$-vertices in ${\cal N}=1$ supersymmetric theories,''
  Nucl.\ Phys.\ B {\bf 909} (2016) 316.
  %doi:10.1016/j.nuclphysb.2016.05.011
  %[arXiv:1603.04801 [hep-th]].
  %%CITATION = doi:10.1016/j.nuclphysb.2016.05.011;%%

%\cite{Kazantsev:2017fdc}
\bibitem{Kazantsev:2017fdc}
  A.~E.~Kazantsev, M.~B.~Skoptsov and K.~V.~Stepanyantz,
  %``One-loop polarization operator of the quantum gauge superfield for ${\cal N}=1$ SYM regularized by higher derivatives,''
  Mod.\ Phys.\ Lett.\ A {\bf 32} (2017) no.36,  1750194.
  %doi:10.1142/S0217732317501942
  %[arXiv:1709.08575 [hep-th]].
  %%CITATION = doi:10.1142/S0217732317501942;%%

%\cite{Kataev:2017qvk}
\bibitem{Kataev:2017qvk}
  A.~L.~Kataev, A.~E.~Kazantsev and K.~V.~Stepanyantz,
  %``The Adler $D$-function for ${\cal N}=1$ SQCD regularized by higher covariant derivatives in the three-loop approximation,''
  Nucl.\ Phys.\ B {\bf 926} (2018) 295.
  %doi:10.1016/j.nuclphysb.2017.11.009
  %[arXiv:1710.03941 [hep-th]].
  %%CITATION = doi:10.1016/j.nuclphysb.2017.11.009;%%

%\cite{Stepanyantz:2017sqg}
\bibitem{Stepanyantz:2017sqg}
  K.~V.~Stepanyantz,
  ``Structure of quantum corrections in ${\cal N}=1$ supersymmetric gauge theories,''
  arXiv:1711.09194 [hep-th].
  %%CITATION = ARXIV:1711.09194;%%

%\cite{Shakhmanov:2017soc}
\bibitem{Shakhmanov:2017soc}
  V.~Y.~Shakhmanov and K.~V.~Stepanyantz,
  %``Three-loop NSVZ relation for terms quartic in the Yukawa couplings with the higher covariant derivative regularization,''
  Nucl.\ Phys.\ B {\bf 920} (2017) 345.
  %doi:10.1016/j.nuclphysb.2017.04.017
  %[arXiv:1703.10569 [hep-th]].
  %%CITATION = doi:10.1016/j.nuclphysb.2017.04.017;%%

%\cite{Kazantsev:2018nbl}
\bibitem{Kazantsev:2018nbl}
  A.~E.~Kazantsev, V.~Y.~Shakhmanov and K.~V.~Stepanyantz,
  %``New form of the exact NSVZ $\beta$-function: the three-loop verification for terms containing Yukawa couplings,''
  JHEP {\bf 1804} (2018) 130.
  %doi:10.1007/JHEP04(2018)130
  %[arXiv:1803.06612 [hep-th]].
  %%CITATION = doi:10.1007/JHEP04(2018)130;%%

%\cite{Juer:1982fb}
\bibitem{Juer:1982fb}
  J.~W.~Juer and D.~Storey,
  %``Nonlinear Renormalization in Superfield Gauge Theories,''
  Phys.\ Lett.\  {\bf 119B} (1982) 125.
  %doi:10.1016/0370-2693(82)90259-3
  %%CITATION = doi:10.1016/0370-2693(82)90259-3;%%

%\cite{Juer:1982mp}
\bibitem{Juer:1982mp}
  J.~W.~Juer and D.~Storey,
  %``One Loop Renormalization of Superfield {Yang-Mills} Theories,''
  Nucl.\ Phys.\ B {\bf 216} (1983) 185.
  %doi:10.1016/0550-3213(83)90491-1
  %%CITATION = doi:10.1016/0550-3213(83)90491-1;%%

%\cite{Piguet:1981fb}
\bibitem{Piguet:1981fb}
  O.~Piguet and K.~Sibold,
  %``Renormalization of $N=1$ Supersymmetrical {Yang-Mills} Theories. 1. The Classical Theory,''
  Nucl.\ Phys.\ B {\bf 197} (1982) 257.
  %doi:10.1016/0550-3213(82)90291-7
  %%CITATION = doi:10.1016/0550-3213(82)90291-7;%%

%\cite{Piguet:1981hh}
\bibitem{Piguet:1981hh}
  O.~Piguet and K.~Sibold,
  %``Renormalization of $N=1$ Supersymmetrical {Yang-Mills} Theories. 2. The Radiative Corrections,''
  Nucl.\ Phys.\ B {\bf 197} (1982) 272.
  %doi:10.1016/0550-3213(82)90292-9
  %%CITATION = doi:10.1016/0550-3213(82)90292-9;%%

%\cite{Piguet:1981mu}
\bibitem{Piguet:1981mu}
  O.~Piguet and K.~Sibold,
  %``The Supercurrent in $N=1$ Supersymmetrical {Yang-Mills} Theories. 1. The Classical Case,''
  Nucl.\ Phys.\ B {\bf 196} (1982) 428.
  %doi:10.1016/0550-3213(82)90499-0
  %%CITATION = doi:10.1016/0550-3213(82)90499-0;%%

%\cite{Piguet:1984mv}
\bibitem{Piguet:1984mv}
  O.~Piguet and K.~Sibold,
  %``Gauge Independence in $N=1$ Supersymmetric {Yang-Mills} Theories,''
  Nucl.\ Phys.\ B {\bf 248} (1984) 301.
  %doi:10.1016/0550-3213(84)90599-6
  %%CITATION = doi:10.1016/0550-3213(84)90599-6;%%

%\cite{Shakhmanov:2017wji}
\bibitem{Shakhmanov:2017wji}
  V.~Y.~Shakhmanov and K.~V.~Stepanyantz,
  %``New form of the NSVZ relation at the two-loop level,''
  Phys.\ Lett.\ B {\bf 776} (2018) 417.
  %doi:10.1016/j.physletb.2017.12.005
  %[arXiv:1711.03899 [hep-th]].
  %%CITATION = doi:10.1016/j.physletb.2017.12.005;%%

%\cite{Soloshenko:2003nc}
\bibitem{Soloshenko:2003nc}
  A.~A.~Soloshenko and K.~V.~Stepanyantz,
  %``Three loop beta function for N=1 supersymmetric electrodynamics, regularized by higher derivatives,''
  Theor.\ Math.\ Phys.\  {\bf 140} (2004) 1264
   [Teor.\ Mat.\ Fiz.\  {\bf 140} (2004) 430].
  %%CITATION = HEP-TH/0304083;%%

%\cite{Smilga:2004zr}
\bibitem{Smilga:2004zr}
  A.~V.~Smilga and A.~Vainshtein,
  %``Background field calculations and nonrenormalization theorems in 4-D supersymmetric gauge theories and their low-dimensional descendants,''
  Nucl.\ Phys.\ B {\bf 704} (2005) 445.
  %%CITATION = HEP-TH/0405142;%%

%\cite{Aleshin:2015qqc}
\bibitem{Aleshin:2015qqc}
  S.~S.~Aleshin, A.~L.~Kataev and K.~V.~Stepanyantz,
  %``Structure of three-loop contributions to the beta-function of N=1 SQED with N_f flavors, regularized by the dimensional reduction,''
  Pisma Zh.\ Eksp.\ Teor.\ Fiz.\  {\bf 130} (2016) 83.
  %[arXiv:1511.05675 [hep-th]].
  %%CITATION = ARXIV:1511.05675;%%

%\cite{Aleshin:2016rrr}
\bibitem{Aleshin:2016rrr}
  S.~S.~Aleshin, I.~O.~Goriachuk, A.~L.~Kataev and K.~V.~Stepanyantz,
  %``The NSVZ scheme for ${\cal N}=1$ SQED with $N_f$ flavors, regularized by the dimensional reduction, in the three-loop approximation,''
  Phys.\ Lett.\ B {\bf 764} (2017) 222.
  %doi:10.1016/j.physletb.2016.11.041
  %[arXiv:1610.08034 [hep-th]].
  %%CITATION = doi:10.1016/j.physletb.2016.11.041;%%

%\cite{Stepanyantz:2011jy}
\bibitem{Stepanyantz:2011jy}
  K.~V.~Stepanyantz,
  %``Derivation of the exact NSVZ $\beta$-function in N=1 SQED, regularized by higher derivatives, by direct summation of Feynman diagrams,''
  Nucl.\ Phys.\ B {\bf 852} (2011) 71.
  %%CITATION = ARXIV:1102.3772;%%

%\cite{Stepanyantz:2014ima}
\bibitem{Stepanyantz:2014ima}
  K.~V.~Stepanyantz,
  %``The NSVZ $\beta$-function and the Schwinger-Dyson equations for $\mathcal{N}=1$ SQED with $N_{f}$ flavors, regularized by higher derivatives,''
  JHEP {\bf 1408} (2014) 096.
  %[arXiv:1404.6717 [hep-th]].
  %%CITATION = ARXIV:1404.6717;%%

%\cite{Adler:1974gd}
\bibitem{Adler:1974gd}
  S.~L.~Adler,
  %``Some Simple Vacuum Polarization Phenomenology: e+ e- ---> Hadrons: The mu - Mesic Atom x-Ray Discrepancy and (g-2) of the Muon,''
  Phys.\ Rev.\ D {\bf 10} (1974) 3714.
  %%CITATION = PHRVA,D10,3714;%%

%\cite{Shifman:2014cya}
\bibitem{Shifman:2014cya}
  M.~Shifman and K.~Stepanyantz,
  %``Exact Adler Function in Supersymmetric QCD,''
  Phys.\ Rev.\ Lett.\  {\bf 114} (2015) 051601.
  %[arXiv:1412.3382 [hep-th]].
  %%CITATION = ARXIV:1412.3382;%%

%\cite{Shifman:2015doa}
\bibitem{Shifman:2015doa}
  M.~Shifman and K.~V.~Stepanyantz,
  %``Derivation of the exact expression for the D function in N=1 SQCD,''
  Phys.\ Rev.\ D {\bf 91} (2015) 105008.
  %[arXiv:1502.06655 [hep-th]].

%\cite{Hisano:1997ua}
\bibitem{Hisano:1997ua}
  J.~Hisano and M.~A.~Shifman,
  %``Exact results for soft supersymmetry breaking parameters in supersymmetric gauge theories,''
  Phys.\ Rev.\ D {\bf 56} (1997) 5475.
  %doi:10.1103/PhysRevD.56.5475
  %[hep-ph/9705417].
  %%CITATION = doi:10.1103/PhysRevD.56.5475;%%

%\cite{Jack:1997pa}
\bibitem{Jack:1997pa}
  I.~Jack and D.~R.~T.~Jones,
  %``The Gaugino Beta function,''
  Phys.\ Lett.\ B {\bf 415} (1997) 383.
  %doi:10.1016/S0370-2693(97)01277-X
  %[hep-ph/9709364].
  %%CITATION = doi:10.1016/S0370-2693(97)01277-X;%%

%\cite{Avdeev:1997vx}
\bibitem{Avdeev:1997vx}
  L.~V.~Avdeev, D.~I.~Kazakov and I.~N.~Kondrashuk,
  %``Renormalizations in softly broken SUSY gauge theories,''
  Nucl.\ Phys.\ B {\bf 510} (1998) 289.
  %doi:10.1016/S0550-3213(98)81015-8, 10.1016/S0550-3213(97)00706-2
  %[hep-ph/9709397].
  %%CITATION = doi:10.1016/S0550-3213(98)81015-8, 10.1016/S0550-3213(97)00706-2;%%

%\cite{Nartsev:2016nym}
\bibitem{Nartsev:2016nym}
  I.~V.~Nartsev and K.~V.~Stepanyantz,
  %``Exact renormalization of the photino mass in softly broken $ \mathcal{N} $ = 1 SQED with N$_{f}$ flavors regularized by higher derivatives,''
  JHEP {\bf 1704} (2017) 047.
  %doi:10.1007/JHEP04(2017)047
  %[arXiv:1610.01280 [hep-th]].
  %%CITATION = doi:10.1007/JHEP04(2017)047;%%

%\cite{Nartsev:2016mvn}
\bibitem{Nartsev:2016mvn}
  I.~V.~Nartsev and K.~V.~Stepanyantz,
  %``NSVZ-like scheme for the photino mass in softly broken ${\cal N}=1$ SQED regularized by higher derivatives,''
  JETP Lett.\  {\bf 105} (2017) no.2,  69.
  %doi:10.1134/S0021364017020059
  %[arXiv:1611.09091 [hep-th]].
  %%CITATION = doi:10.1134/S0021364017020059;%%

%\cite{Aleshin:2016yvj}
\bibitem{Aleshin:2016yvj}
  S.~S.~Aleshin, A.~E.~Kazantsev, M.~B.~Skoptsov and K.~V.~Stepanyantz,
  %``One-loop divergences in non-Abelian supersymmetric theories regularized by BRST-invariant version of the higher derivative regularization,''
  JHEP {\bf 1605} (2016) 014.
  %doi:10.1007/JHEP05(2016)014
  %[arXiv:1603.04347 [hep-th]].
  %%CITATION = doi:10.1007/JHEP05(2016)014;%%

%\cite{Pimenov:2009hv}
\bibitem{Pimenov:2009hv}
  A.~B.~Pimenov, E.~S.~Shevtsova and K.~V.~Stepanyantz,
  %``Calculation of two-loop beta-function for general N=1 supersymmetric Yang--Mills theory with the higher covariant derivative regularization,''
  Phys.\ Lett.\ B {\bf 686} (2010) 293.
  %%CITATION = ARXIV:0912.5191;%%

\bibitem{Stepanyantz_MIAN}
  K.~V.~Stepanyantz,
  %``Higher covariant derivative regularization for calculations in supersymmetric theories,''
  Proceedings of the Steklov Institute of Mathematics\  {\bf 272} (2011) 256.

%\cite{Stepanyantz:2011bz}
\bibitem{Stepanyantz:2011bz}
  K.~V.~Stepanyantz,
  ``Factorization of integrals defining the two-loop $\beta$-function for the general renormalizable N=1 SYM theory, regularized by the higher covariant derivatives, into integrals of double total derivatives,''
  arXiv:1108.1491 [hep-th].
  %%CITATION = ARXIV:1108.1491;%%

%\cite{Stepanyantz:2012zz}
\bibitem{Stepanyantz:2012zz}
  K.~V.~Stepanyantz,
  %``Derivation of the exact NSVZ beta-function in N=1 SQED regularized by higher derivatives by summation of Feynman diagrams,''
  J.\ Phys.\ Conf.\ Ser.\  {\bf 343} (2012) 012115.
  %%CITATION = 00462,343,012115;%%

%\cite{Stepanyantz:2012us}
\bibitem{Stepanyantz:2012us}
  K.~V.~Stepanyantz,
  %``Multiloop calculations in supersymmetric theories with the higher covariant derivative regularization,''
  J.\ Phys.\ Conf.\ Ser.\  {\bf 368} (2012) 012052.
  %%CITATION = ARXIV:1203.5525;%%

%\cite{Kazantsev:2014yna}
\bibitem{Kazantsev:2014yna}
  A.~E.~Kazantsev and K.~V.~Stepanyantz,
  %``Relation between two-point Green functions of ${\cal N}=1$ SQED with $N_f$ flavors, regularized by higher derivatives, in the three-loop approximation,''
  J.\ Exp.\ Theor.\ Phys. {\bf 120} (2015) 618 [Zh.\ Eksp.\ Teor.\ Fiz. {\bf 147} (2015) 714].
  %[arXiv:1410.1133 [hep-th]].
  %%CITATION = ARXIV:1410.1133;%%

%\cite{Jack:2005ij}
\bibitem{Jack:2005ij}
  I.~Jack, D.~R.~T.~Jones and L.~A.~Worthy,
  %``Renormalisation of supersymmetric gauge theory in the uneliminated component formalism,''
  Phys.\ Rev.\ D {\bf 72} (2005) 107701.
  %doi:10.1103/PhysRevD.72.107701
  %[hep-th/0509089].
  %%CITATION = doi:10.1103/PhysRevD.72.107701;%%

%\cite{Faddeev:1980be}
\bibitem{Faddeev:1980be}
  L.~D.~Faddeev and A.~A.~Slavnov,
``Gauge Fields. Introduction To Quantum Theory,'' Nauka, Moscow,
1978 and
  Front.\ Phys.\  {\bf 50} (1980) 1
   [Front.\ Phys.\  {\bf 83} (1990) 1].
  %%CITATION = FRPHA,50,1;%%

%\cite{Chetyrkin:1980sa}
\bibitem{Chetyrkin:1980sa}
  K.~G.~Chetyrkin, A.~L.~Kataev and F.~V.~Tkachov,
  ``Computation of the $\alpha^2_s$ Correction Sigma-t ($e^+ e^- \to$ Hadrons) in {QCD},''
  IYaI-P-0170.
  %%CITATION = IYAI-P-0170;%%

%\cite{Pronin:1997eb}
\bibitem{Pronin:1997eb}
  P.~I.~Pronin and K.~Stepanyantz,
  %``One loop counterterms for higher derivative regularized Lagrangians,''
  Phys.\ Lett.\ B {\bf 414} (1997) 117.
  %doi:10.1016/S0370-2693(97)01147-7
  %[hep-th/9707008].
  %%CITATION = doi:10.1016/S0370-2693(97)01147-7;%%

%\cite{Soloshenko:2002np}
\bibitem{Soloshenko:2002np}
  A.~Soloshenko and K.~Stepanyantz,
  ``Two loop renormalization of N=1 supersymmetric electrodynamics, regularized by higher derivatives,''
  hep-th/0203118.
  %%CITATION = HEP-TH/0203118;%%


\end{thebibliography}
\end{document}